\documentclass[11pt,a4paper,twoside]{article}
\pdfoutput=1
\usepackage[pdftex]{graphicx}
\usepackage[latin1]{inputenc}

\usepackage{thumbpdf}

\usepackage{bbm}
\usepackage{amsmath,amsfonts,amssymb}
\usepackage{mathrsfs}
\usepackage[footnotesize]{caption}

\allowdisplaybreaks[1]

\begin{document}

\begin{titlepage}

\vspace*{-15mm}
\begin{flushright}
MPP-2009-26\\
\end{flushright}
\vspace*{0.7cm}

\begin{center}
{
\bf\LARGE
New GUT predictions for quark and lepton \\[1mm] 
mass ratios confronted with phenomenology
}
\\[8mm]
S.~Antusch$^{\star}$
\footnote{E-mail: \texttt{antusch@mppmu.mpg.de}}, 
M.~Spinrath$^{\star}$
\footnote{E-mail: \texttt{spinrath@mppmu.mpg.de}},
\\[1mm]

\end{center}
\vspace*{0.50cm}
\centerline{$^{\star}$ \it 
Max-Planck-Institut f\"ur Physik (Werner-Heisenberg-Institut),}
\centerline{\it 
F\"ohringer Ring 6, D-80805 M\"unchen, Germany}
\vspace*{1.20cm}

\begin{abstract}

\noindent 
Group theoretical factors from GUT symmetry breaking can lead to predictions for the ratios of quark and lepton masses (or Yukawa couplings) at the unification scale. Due to supersymmetric (SUSY) threshold corrections the viability of such predictions can depend strongly on the SUSY parameters. For three common minimal SUSY breaking scenarios with anomaly, gauge and gravity mediation we investigate which GUT scale ratios $m_e/m_d$, $m_\mu/m_s$, $y_\tau/y_b$ and $y_t/y_b$ are allowed when phenomenological constraints from electroweak precision observables, $B$ physics, $(g-2)_\mu$, mass limits on sparticles from direct searches as well as, optionally, dark matter constraints are taken into account. We derive possible new predictions for the GUT scale mass ratios and compare them with the phenomenologically allowed ranges. We find that new GUT scale predictions such as $m_\mu/m_s = 9/2$ or $6$ and $y_\tau/y_b = 3/2$ or $2$ are often favoured compared to the ubiquitous relations $m_\mu/m_s = 3$ or $y_\tau/y_b =1$. They are viable for characteristic SUSY scenarios, testable at the CERN LHC and future colliders.
\end{abstract}

\end{titlepage}

\newpage
\setcounter{footnote}{0}

\section{Introduction}

The origin of the observed pattern of fermion masses and mixings is one of the great open questions in particle physics. In this context, it is particularly challenging to explain the strong hierarchy among the masses of the three families of quarks and charged leptons, as well as the strong suppression of the neutrino masses and the fact that quark mixings are small whereas there is large mixing between mass and flavour eigenstates in the lepton sector. One possibility to address the fermion mass hierarchy is to introduce family symmetries which allow Yukawa couplings, in particular for the first and second generation, only via higher-dimensional effective operators leading to a certain suppression compared to apparently natural ${\cal O}(1)$ values.

An interesting observation in this context is that in supersymmetric theories with large (or medium) $\tan \beta$, the Yukawa couplings of each of the three generations of fermions in the up-quark, down-quark and charged lepton sector are of similar order of magnitude. This observation can have an explanation in Grand Unified Theories (GUTs), where not only the gauge interactions of the Standard Model (SM) emerge from one unified gauge group, but also the quarks and leptons are unified in joint representations. 
In these theories, the Yukawa couplings for different types of fermions of one generation can be generated from common operators involving the GUT representations. After GUT symmetry breaking the resulting Yukawa couplings typically have similar values. Furthermore, depending on the specific operator, the group theoretical Clebsch factors from GUT symmetry breaking can lead to predictions for the ratios between the Yukawa couplings (see, e.g., \cite{GJ}). Such relations, after evolving them from the GUT scale to low energies via their renormalisation group equations and including threshold effects \cite{SUSYthresholds,Carena:1994bv,Hempfling:1993kv,Blazek}, can be compared to experimental results for the quark masses and provide crucial tests of unified models of fermion masses and mixings.

When testing the predictions of supersymmetric (SUSY) GUTs for quark and lepton mass ratios\footnote{We note that when we refer to fermion masses at the GUT scale, what we mean is simply the Yukawa
coupling multiplied by the low-energy value of the corresponding Higgs vev.} 
 $m_e/m_d$, $m_\mu/m_s$, $y_\tau/y_b$ and $y_t/y_b$ the $\tan \beta$-enhanced SUSY threshold corrections are of particular importance, as has been emphasized recently in \cite{Ross:2007az,Antusch:2008tf}. Including the $\tan \beta$-enhanced SUSY threshold corrections the allowed values of the GUT scale Yukawa couplings and their ratios have been calculated in \cite{Antusch:2008tf} for example ranges of low energy SUSY parameters. It has also been pointed out in \cite{Antusch:2008tf} that the presence of SUSY threshold corrections can open up new possibilities for GUT model building. 

The goal of this study is to investigate which new predictions for Yukawa coupling ratios at the GUT scale can arise in unified models and whether they can be realised in common scenarios of SUSY breaking in a phenomenologically acceptable way. For this purpose we consider the three SUSY breaking schemes mAMSB \cite{AMSB}, mGMSB \cite{Ambrosanio:1997rv} and CMSSM \cite{mSUGRA, mSUGRArev} which provide boundary conditions for the soft SUSY breaking parameters at high energies. As phenomenological constraints we will apply observables from electroweak precision data, $B$ physics, $(g-2)_\mu$, mass limits on sparticles from direct searches as well as, optionally, dark matter constraints. 

The paper is organised as follows: In section 2 we investigate possible predictions for GUT scale ratios of quark and lepton masses (or Yukawa couplings) in unified theories. In section 3 we calculate the allowed GUT scale ranges for the ratios $m_e/m_d$, $m_\mu/m_s$, $y_\tau/y_b$ and $y_t/y_b$ in mAMSB, mGMSB and CMSSM taking phenomenological constraints into account. The results of section 3 are independent of the details of the underlying GUT theory. Section 4 contains the comparison of possible theory predictions for the GUT scale ratios with the phenomenologically allowed ranges. In section 5 we summarise our main results and conclude.

\section{GUT predictions for quark and lepton mass relations} \label{sec:Relations}

In the following, we will consider unified theories where the fermions of the SM are unified in representations of the unifying gauge group. We will focus on supersymmetric $SO(10)$ GUTs where the symmetry breaking to the MSSM proceeds via the $SU(5)$ or Pati-Salam (PS) breaking chain at the GUT scale $M_{GUT}\approx 2 \times 10^{16}$~GeV. We will perform our analysis at the stage of $SU(5)$ or PS unification for simplicity, however our choice of possible GUT Higgs representations will be motivated by the embedding into $SO(10)$ GUTs.     
Within such unified theories the Yukawa couplings emerge from operators involving the joint fermion representations as well as Higgs fields in GUT representations of which one has to contain an electroweak  Higgs. Each such operator thus in general generates Yukawa couplings for different types of fermions, for example for down-type quarks as well as for charged leptons, which are related to each other by the group theoretical Clebsch factors from the breaking of the GUT symmetry to the MSSM. 

\subsection{Conditions for the appearance of predictions} \label{sec:Conditions}

Let us now clarify under which conditions such relations lead to observable predictions for quark and lepton masses. One condition, which results in simple relations between entries of the Yukawa matrices and the quark and charged lepton masses is that the Yukawa matrices in the flavour basis are hierarchical and dominated by the diagonal elements. This situation is approximately realised in many approaches to unified model building, but only regarding the second and third generation. Then, the masses of the second generation of quarks and charged leptons are related to the (2,2)-entries of the Yukawa matrices and the masses of the third generation to the (3,3)-entries. For the masses of the first generation of fermions the condition is often violated and the relation to the elements of the Yukawa matrices often depends on additional assumptions, e.g.\ if there is a texture zero in the (1,1)-entry of the Yukawa matrices (see e.g.\ \cite{GST}). We will therefore focus mainly on the second and third generation in our analysis. 
The second condition is that there is one operator which dominates the relevant element of the Yukawa matrices. This requirement is necessary because if, for instance, two operators would contribute with similar strength, the resulting prediction would be an intermediate value. 
In the following we will therefore assume that these two conditions are satisfied to good approximation.

\subsection{Examples: Bottom-tau unification and Georgi-Jarlskog relations}
There are two examples of quark and lepton mass relations at the GUT scale which are ubiquitous in many classes of unified models of flavour. These are third family Yukawa unification (or $b$-$\tau$ unification) and the so-called Georgi-Jarlskog (GJ) relations \cite{GJ} (i.e.\ $m_\mu/m_s=3$ and $m_e/m_d=1/3$). Let us briefly review them in the context of $SU(5)$ GUTs to give an explicit example: 
In $SU(5)$ GUTs, the $SU(2)_L$ singlet down-type quarks (in three colours) as well as the $SU(2)_L$ doublet leptons of the $i$-th generation are contained in the fundamental representation $F_5^i$ while $SU(2)_L$ doublet quarks as well as singlet up-type quarks and charged leptons are contained in a ten-dimensional matter representation $\bar{F}_{10}^i$ (see section \ref{sec:SU5} for more details). If the Yukawa matrix (3,3)-entries for down-type quarks and charged leptons are generated by an operator of the form $F_5^3 \bar{F}_{10}^3 H_5$ where the five-dimensional $H_5$ contains an electroweak Higgs $SU(2)_L$ doublet, then it is easy to see that the resulting prediction is $y_b/y_t = 1$, i.e.\ approximate $b$-$\tau$ unification. On the other hand, if the relevant (2,2)-entry of the Yukawa matrices is generated by the operator $F_5^2 \bar{F}_{10}^2 H_{45}$ with electroweak Higgs fields contained in the $45$-dimensional representation $H_{45}$ then $m_\mu / m_s = -3$ is predicted. This can be understood from the fact that the $45$-dimensional representation is traceless and the factor of $-3$ for the charged leptons thus has to compensate the colour factor of $3$ for the quarks. 

In addition to $b$-$\tau$ unification and the GJ relations there are various alternative relations between quark and lepton masses which can emerge from higher-dimensional operators in unified theories, as we will now discuss.

\subsection{New GUT predictions}
When the conditions specified in section \ref{sec:Conditions} are satisfied, 
the predicted relations between quark and lepton masses at the GUT scale depend on the specific operator which dominates the relevant entries of the Yukawa matrices. The simplest types of operators in this context are the renormalisable ones, for example the operators mentioned above which lead to $b$-$\tau$ unification and the GJ  relation for the second generation. The different predictions result from different Higgs representations which can contain the electroweak Higgs(es). Here the general procedure to obtain the possible predictions for the Yukawa coupling ratios is as follows: The operators include two matter and one Higgs field. For the matter fields we take
the common matter representations of the unified theories. By fixing
two of the three fields, the possible representations of the Higgs
field are fixed by the condition that the operator has to be a gauge
singlet after contracting all gauge indices and that the Higgs field
has to include the usual SM (MSSM) Higgs(es). Explicit expressions for the
matter fields and the Higgs vacuum expectation values (vevs) will be given later.

\begin{figure}[hbt]
 \centering
 \includegraphics[scale=0.7]{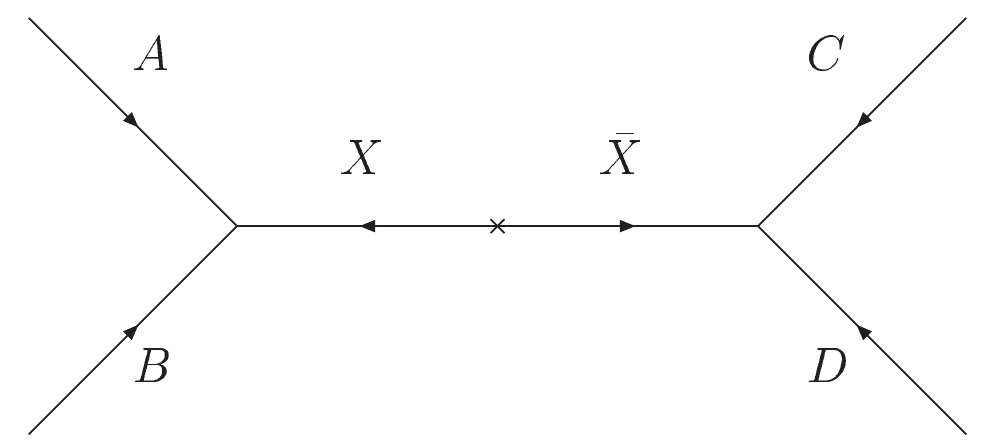}
 \caption{Supergraph with heavy messenger fields $X$ and $\bar{X}$. When the messenger fields are effectively integrated out of the theory below their mass scales, higher-dimensional operators are generated which can lead to GUT relations between quark and lepton masses. \label{fig:Messenger}}
\end{figure}

New possibilities, in addition to the ratios $1$ and $-3$ can arise in particular when effective, higher-dimensional operators are taken into account. As has been discussed in the introduction, in many unified flavour models using family symmetries to explain the observed fermion mass hierarchy, the renormalisable (dimension-four) operators are forbidden by symmetry, and the Yukawa couplings are generated from higher-dimensional operators in the effective theory limit. 
These non-renormalisable operators are typically generated from integrating out messenger fields $X$
and $\bar{X}$ (c.f.\ figure \ref{fig:Messenger}).
The fields $A$, $B$, $C$ and $D$ can be either a matter field
or a Higgs field. In total the effective operator has to contain two
matter fields, one Higgs field which breaks electroweak symmetry
and one Higgs field with a GUT scale vev. The latter must only 
break the unified gauge symmetry but not the electroweak symmetry. 
At low energies, the Yukawa operators of the SM (MSSM) are realised
with some of the Yukawa couplings related to each other due to the 
underlying unified group structure.

In our study we will restrict ourselves to messenger fields and GUT scale
Higgs fields which are included in the common $SO(10)$ representations, i.e.\
$\mathbf{10}$, $\mathbf{16}$, $\mathbf{45}$,
$\mathbf{54}$, $\mathbf{120}$, $\mathbf{126}$ and $\mathbf{210}$ of
$SO(10)$. With these restrictions, we cover the cases of most GUT models based on 
$SO(10)$ broken to the SM gauge group via PS or $SU(5)$ using the above listed 
Higgs representations (see e.g.\ \cite{Slansky}). 
In the next subsections we will derive the results for the cases of (SO(10) broken
to the SM via) $SU(5)$ or PS. A summary of the results is contained in tables
\ref{tab:SU5relations} and \ref{tab:PSrelations}.

\subsubsection{Predictions from $\boldsymbol{SU(5)}$ unification} \label{sec:SU5}

\begin{table}
\begin{center}
\begin{tabular}{|c|c|c|c|} \hline
($A$, $B$) & ($C$, $D$) & $X$ & $y_e/y_d$ \\ \hline
($\mathbf{5}_F$, $\mathbf{\bar{10}}_F$) & ($\mathbf{5}_H$, $\mathbf{24}_H$) & $\mathbf{5}$ & 1 \\
($\mathbf{5}_F$, $\mathbf{\bar{10}}_F$) & ($\mathbf{5}_H$, $\mathbf{24}_H$) & $\mathbf{45}$ & -3 \\
($\mathbf{5}_F$, $\mathbf{\bar{10}}_F$) & ($\mathbf{5}_H$, $\mathbf{75}_H$) & $\mathbf{45}$ & -3 \\
($\mathbf{5}_F$, $\mathbf{\bar{10}}_F$) & ($\mathbf{45}_H$, $\mathbf{24}_H$) & $\mathbf{5}$ & 1 \\
($\mathbf{5}_F$, $\mathbf{\bar{10}}_F$) & ($\mathbf{45}_H$, $\mathbf{24}_H$) & $\mathbf{45}_1$ & -3 \\
($\mathbf{5}_F$, $\mathbf{\bar{10}}_F$) & ($\mathbf{45}_H$, $\mathbf{24}_H$) & $\mathbf{45}_2$ & - \\
($\mathbf{5}_F$, $\mathbf{\bar{10}}_F$) & ($\mathbf{45}_H$, $\mathbf{75}_H$) & $\mathbf{5}$ & 1 \\
($\mathbf{5}_F$, $\mathbf{\bar{10}}_F$) & ($\mathbf{45}_H$, $\mathbf{75}_H$) & $\mathbf{45}_1$ & -3 \\
($\mathbf{5}_F$, $\mathbf{\bar{10}}_F$) & ($\mathbf{45}_H$, $\mathbf{75}_H$) & $\mathbf{45}_2$ & - \\
($\mathbf{5}_F$, $\mathbf{5}_H$) & ($\mathbf{\bar{10}}_F$, $\mathbf{24}_H$) & $\mathbf{\bar{10}}$ & 6 \\
($\mathbf{5}_F$, $\mathbf{5}_H$) & ($\mathbf{\bar{10}}_F$, $\mathbf{24}_H$) & $\mathbf{\bar{15}}$ & 0 \\
($\mathbf{5}_F$, $\mathbf{5}_H$) & ($\mathbf{\bar{10}}_F$, $\mathbf{75}_H$) & $\mathbf{\bar{10}}$ & -3 \\
($\mathbf{5}_F$, $\mathbf{45}_H$) & ($\mathbf{\bar{10}}_F$, $\mathbf{24}_H$) & $\mathbf{\bar{10}}$ & -18 \\
($\mathbf{5}_F$, $\mathbf{45}_H$) & ($\mathbf{\bar{10}}_F$, $\mathbf{24}_H$) & $\mathbf{\bar{40}}$ & 0 \\
($\mathbf{5}_F$, $\mathbf{45}_H$) & ($\mathbf{\bar{10}}_F$, $\mathbf{75}_H$) & $\mathbf{\bar{10}}$ & 9 \\
($\mathbf{5}_F$, $\mathbf{45}_H$) & ($\mathbf{\bar{10}}_F$, $\mathbf{75}_H$) & $\mathbf{\bar{40}}$ & 0 \\
($\mathbf{5}_F$, $\mathbf{24}_H$) & ($\mathbf{\bar{10}}_F$, $\mathbf{5}_H$) & $\mathbf{\bar{5}}$ & -3/2 \\
($\mathbf{5}_F$, $\mathbf{24}_H$) & ($\mathbf{\bar{10}}_F$, $\mathbf{5}_H$) & $\mathbf{\bar{45}}$ & 3/2 \\
($\mathbf{5}_F$, $\mathbf{75}_H$) & ($\mathbf{\bar{10}}_F$, $\mathbf{5}_H$) & $\mathbf{\bar{45}}$ & -3 \\
($\mathbf{5}_F$, $\mathbf{24}_H$) & ($\mathbf{\bar{10}}_F$, $\mathbf{45}_H$) & $\mathbf{\bar{5}}$ & 9/2 \\
($\mathbf{5}_F$, $\mathbf{24}_H$) & ($\mathbf{\bar{10}}_F$, $\mathbf{45}_H$) & $\mathbf{\bar{45}}$ & -1/2 \\
($\mathbf{5}_F$, $\mathbf{75}_H$) & ($\mathbf{\bar{10}}_F$, $\mathbf{45}_H$) & $\mathbf{\bar{45}}$ & 1 \\
($\mathbf{5}_F$, $\mathbf{75}_H$) & ($\mathbf{\bar{10}}_F$, $\mathbf{45}_H$) & $\mathbf{\bar{50}}$ & 0 \\
\hline
\end{tabular}
\end{center}
\caption{
Dimension-five operators within $SU(5)$ unification and resulting predictions for the GUT scale ratios $y_e/y_d$, where $e$ and $d$ stand for any charged
lepton and down-type quark of the same generation. $A, B, C, D$ and $X$ correspond to the fields in the supergraph for Yukawa couplings in figure \ref{fig:Messenger} which generates the dimension-five operator after integrating out the heavy messenger field. 
If the messenger representation $X$ has an index, there is more than one way to combine the fields $A$ and $B$ or $C$ and $D$ to form this representation leading to different predicted ratios $y_e/y_d$. 
A dash means that $y_d$ is zero. 
At the stage of $SU(5)$ unification the dimension-five operators predict no relation to the up-type quark or neutrino Yukawa couplings.
\label{tab:SU5messenger}}
\end{table}

\begin{table}
\begin{center}
\begin{tabular}{|c|c|}\hline
Operator dimension &	$y_e/y_d$ 	\\ \hline
\hline 4	&	1	\\
	&	-3	\\
\hline 5	&	-1/2	\\
	&	1	\\
	&	$\pm$3/2	\\
	&	-3	\\
	&	9/2	\\
	&	6	\\
	&	9	\\
	&	-18	\\
\hline
\end{tabular}
\end{center}
\caption{
Summary of possible $SU(5)$ predictions for the GUT scale ratios $y_e/y_d$, where $e$ and $d$ stand for any charged lepton and down-type quark of the same generation. 
\label{tab:SU5relations}}
\end{table}

As mentioned above, we perform our analysis at the stage of $SU(5)$ or PS unification for simplicity, however 
we have in mind a possible embedding into $SO(10)$ GUTs. 
In GUTs based on the unifying gauge group $SU(5)$ the fermions of the SM are embedded in the GUT
representations $\mathbf{5}_F$ and $\mathbf{\bar{10}}_F$ in the following way:  
The $SU(2)_L$ singlet down-type quarks (in three colours) as well as the $SU(2)_L$ doublet leptons of the $i$-th generation are contained in the fundamental representation $F_5^i$ as
\begin{eqnarray}
 F_5^i &=& \mathbf{5}^i_F = \begin{pmatrix}
           d_R^R & d_R^B & d_R^G & e_L^c &-\nu_L^c
           \end{pmatrix}^i
\end{eqnarray}
while $SU(2)_L$ doublet quarks as well as singlet up-type quarks and charged leptons are contained in a ten-dimensional matter representation $\bar{F}_{10}^i$ as
\begin{eqnarray}
 \bar{F}_{10}^i &=& \mathbf{\bar{10}}^i_F = \frac{1}{\sqrt{2}}
            \begin{pmatrix}
            0 & -u_R^G & u_R^B & -u_L^{c R} & -d_L^{c R} \\
            u_R^G & 0 & -u_R^R & -u_L^{c B} & -d_L^{c B} \\
            -u_R^B & u_R^R & 0 & -u_L^{c G} & -d_L^{c G} \\
            u_L^{c R} & u_L^{c B} & u_L^{c G} & 0 & -e_R \\
            d_L^{c R} & d_L^{c B} & d_L^{c G} & e_R & 0
            \end{pmatrix}^i,
\end{eqnarray}
where $i=1,2,3$ is the family index, the upper indices $R$, $B$ and
$G$ denote color, $c$ stands for charge conjugation and the lower
index $L$($R$) stands for $SU(2)_L$ doublet ($SU(2)_L$ singlet). The $F_5^i$ and $\bar{F}_{10}^i$, plus an extra SM singlet, form the matter representations $\mathbf{16}^i$ of $SO(10)$.  

The commonly used GUT representations which contain the Higgs fields are $\mathbf{5}_H$, $\mathbf{24}_H$ and
$\mathbf{45}_H$. Their notation and vevs are specified as
\begin{eqnarray}
 \left(H_5\right)^a = \mathbf{5}_H &,& \;
   \langle \left(H_5\right)^5 \rangle = v_5 \;\:,\\
\label{Eq:H24vev} \left(H_{24}\right)^a_b = \mathbf{24}_H &,& \;
   \langle \left(H_{24}\right)^a_a \rangle =
      v_{24} (2 \delta_{a\alpha} -3 \delta_{a\beta})\;\:,\\
 \left(H_{45}\right)^{ab}_c = -\left(H_{45}\right)^{ba}_c
   = \mathbf{45}_H &,& \; \langle \left(H_{45}\right)^{i5}_j \rangle =
          v_{45} \left(\delta^i_j - 4 \delta^{i4} \delta_{j4}\right)\;\:,
\end{eqnarray}
where $a,b=1,\ldots,5$, $\alpha=1,2,3$, $\beta=4,5$ and
$i,j=1,\ldots,4$. The vevs $v_5$ and $v_{45}$ are assumed to be of
the electroweak scale whereas $v_{24}$ is of the order of the GUT
scale. The $\mathbf{24}_H$ breaks $SU(5)$. For the determination of
the vevs of the GUT-breaking Higgs fields we have 
neglected the vevs of the Higgs fields which break the electroweak symmetry (which provides a 
very good approximation).

In addition, we also consider the Higgs representation $\mathbf{75}_H$. 
$\mathbf{24}_H$ and $\mathbf{75}_H$ are the only nontrivial representations
which are included in the common $SO(10)$ representations and have a SM singlet 
component that can obtain a GUT scale vev without breaking the
SM symmetries. We construct the vev of $\mathbf{75}_H$ from the vev of 
$\mathbf{24}_H$, which preserves $SU(3)_C \times SU(2)_L \times
U(1)_Y$.

On dimension four only operators containing $\mathbf{5}_H$ and
$\mathbf{45}_H$ can generate Yukawa couplings. The first one gives
$b$-$\tau$ unification and the latter the GJ
relation mentioned in section \ref{sec:Relations}.
For dimension five we can add an
additional $\mathbf{24}_H$ or a $\mathbf{75}_H$ 
to the dimension-four operators. All possible
combinations of external and messenger fields are listed in
table \ref{tab:SU5messenger}, including the corresponding Yukawa
coupling ratio. If the messenger representation in the table has an
index, there was more than one way to combine the fields $A$ and $B$
or $C$ and $D$ to form this representation.
The resulting relations are listed in table \ref{tab:SU5relations}. Since the operators do not relate the up-type quarks to the down-type quarks or charged leptons, we only present the predicted ratio 
$y_e/y_d$, where $e$ and $d$ stand for any charged
lepton and down-type quark of the same generation.
Higher-dimensional operators involving the Higgs representation $\mathbf{24}_H$ have also been considered in \cite{Duque:2008ah}. The possible Clebsch factor $3/2$ is mentioned there as well, however it has not been postulated as a GUT prediction.  

To illustrate how the relations from dimension five operators are generated,
let us discuss the operator leading to the new prediction $y_e/y_d = 9/2$. 
Using the notation of figure 1 we can assign $A=\mathbf{5}_F$, $B=\mathbf{24}_H$, $C=\mathbf{\bar{10}}_F$ and $D=\mathbf{45}_H$. At the left vertex $\mathbf{5}_F$ and $\mathbf{24}_H$ are combined to a $\mathbf{5}$ to couple to the messenger field $X=\mathbf{\bar 5}$. From the vev of $\mathbf{24}_H$ the down-type quarks
are multiplied by a factor of 2 and the leptons by a factor of -3 (c.f.~Eq.~(\ref{Eq:H24vev})). 
At the right vertex $\mathbf{\bar{10}}_F$ and $\mathbf{45}_H$ are
combined to form a $\mathbf{\bar 5}$. Since $\mathbf{45}_H$ is traceless, this, similar to the GJ relation, leads to an additional relative factor of $-3$ for the down-type quarks compared to the charged leptons. In combination, this gives a relative factor of $9/2$.

\subsubsection{Predictions from Pati-Salam Unification (embedded in $\boldsymbol{SO(10)}$ GUTs)} \label{sec:PS}

\begin{table}
\begin{center}
\begin{tabular}{|c|c|c|c|} \hline
($A$, $B$) & ($C$, $D$) & $X$ & $(y_e/y_d,y_u/y_d,y_\nu/y_u)$ \\ \hline
($(\mathbf{4},\mathbf{2},\mathbf{1})_F$, 
$(\mathbf{1},\mathbf{1},\mathbf{3})_H$) & 
($(\mathbf{\bar{4}},\mathbf{1},\mathbf{\bar{2}})_F$, 
$(\mathbf{\bar{1}},\mathbf{\bar{2}},\mathbf{2})_H$) & 
$(\mathbf{\bar{4}},\mathbf{\bar{2}},\mathbf{\bar{3}})$ & (1,1,1) \\
($(\mathbf{4},\mathbf{2},\mathbf{1})_F$, 
$(\mathbf{\bar{4}},\mathbf{1},\mathbf{\bar{2}})_F$) & 
($(\mathbf{\bar{1}},\mathbf{\bar{2}},\mathbf{2})_H$, 
$(\mathbf{15},\mathbf{1},\mathbf{1})_H$) & 
$(\mathbf{15},\mathbf{\bar{2}},\mathbf{2})$ & (-3,1,-3) \\
($(\mathbf{4},\mathbf{2},\mathbf{1})_F$, 
$(\mathbf{\bar{4}},\mathbf{1},\mathbf{\bar{2}})_F$) & 
($(\mathbf{\bar{1}},\mathbf{\bar{2}},\mathbf{2})_H$, 
$(\mathbf{15},\mathbf{1},\mathbf{3})_H$) & 
$(\mathbf{15},\mathbf{\bar{2}},\mathbf{2})$ & (-3,1,-3) \\
($(\mathbf{4},\mathbf{2},\mathbf{1})_F$, 
$(\mathbf{\bar{4}},\mathbf{1},\mathbf{\bar{2}})_F$) & 
($(\mathbf{15},\mathbf{\bar{2}},\mathbf{2})_H$, 
$(\mathbf{1},\mathbf{1},\mathbf{3})_H$) & 
$(\mathbf{15},\mathbf{\bar{2}},\mathbf{2})$ & (-3,1,-3) \\
($(\mathbf{4},\mathbf{2},\mathbf{1})_F$, 
$(\mathbf{\bar{4}},\mathbf{1},\mathbf{\bar{2}})_F$) & 
($(\mathbf{15},\mathbf{\bar{2}},\mathbf{2})_H$, 
$(\mathbf{15},\mathbf{1},\mathbf{1})_H$) & 
$(\mathbf{1},\mathbf{\bar{2}},\mathbf{2})$ & (1,1,1) \\
($(\mathbf{4},\mathbf{2},\mathbf{1})_F$, 
$(\mathbf{\bar{4}},\mathbf{1},\mathbf{\bar{2}})_F$) & 
($(\mathbf{15},\mathbf{\bar{2}},\mathbf{2})_H$, 
$(\mathbf{15},\mathbf{1},\mathbf{1})_H$) & 
$(\mathbf{15}_1,\mathbf{\bar{2}},\mathbf{2})$ & (-3,1,-3) \\
($(\mathbf{4},\mathbf{2},\mathbf{1})_F$, 
$(\mathbf{\bar{4}},\mathbf{1},\mathbf{\bar{2}})_F$) & 
($(\mathbf{15},\mathbf{\bar{2}},\mathbf{2})_H$, 
$(\mathbf{15},\mathbf{1},\mathbf{1})_H$) & 
$(\mathbf{15}_2,\mathbf{\bar{2}},\mathbf{2})$ & (-3,1,-3) \\
($(\mathbf{4},\mathbf{2},\mathbf{1})_F$, 
$(\mathbf{15},\mathbf{\bar{2}},\mathbf{2})_H$) & 
($(\mathbf{\bar{4}},\mathbf{1},\mathbf{\bar{2}})_F$, 
$(\mathbf{15},\mathbf{1},\mathbf{1})_H$) & 
$(\mathbf{\bar{4}},\mathbf{1},\mathbf{\bar{2}})$ & (9,1,9) \\
($(\mathbf{\bar{4}},\mathbf{1},\mathbf{\bar{2}})_F$, 
$(\mathbf{15},\mathbf{\bar{2}},\mathbf{2})_H$) & 
($(\mathbf{4},\mathbf{2},\mathbf{1})_F$, 
$(\mathbf{15},\mathbf{1},\mathbf{1})_H$) & 
$(\mathbf{4},\mathbf{2},\mathbf{1})$ & (9,1,9) \\
($(\mathbf{4},\mathbf{2},\mathbf{1})_F$, 
$(\mathbf{\bar{4}},\mathbf{1},\mathbf{\bar{2}})_F$) & 
($(\mathbf{15},\mathbf{\bar{2}},\mathbf{2})_H$, 
$(\mathbf{15},\mathbf{1},\mathbf{3})_H$) & 
$(\mathbf{1},\mathbf{\bar{2}},\mathbf{2})$ & (1,1,1) \\
($(\mathbf{4},\mathbf{2},\mathbf{1})_F$, 
$(\mathbf{\bar{4}},\mathbf{1},\mathbf{\bar{2}})_F$) & 
($(\mathbf{15},\mathbf{\bar{2}},\mathbf{2})_H$, 
$(\mathbf{15},\mathbf{1},\mathbf{3})_H$) & 
$(\mathbf{15}_1,\mathbf{\bar{2}},\mathbf{2})$ & (-3,1,-3) \\
($(\mathbf{4},\mathbf{2},\mathbf{1})_F$, 
$(\mathbf{\bar{4}},\mathbf{1},\mathbf{\bar{2}})_F$) & 
($(\mathbf{15},\mathbf{\bar{2}},\mathbf{2})_H$, 
$(\mathbf{15},\mathbf{1},\mathbf{3})_H$) & 
$(\mathbf{15}_2,\mathbf{\bar{2}},\mathbf{2})$ & (-3,1,-3) \\
($(\mathbf{4},\mathbf{2},\mathbf{1})_F$, 
$(\mathbf{15},\mathbf{\bar{2}},\mathbf{2})_H$) & 
($(\mathbf{\bar{4}},\mathbf{1},\mathbf{\bar{2}})_F$, 
$(\mathbf{15},\mathbf{1},\mathbf{3})_H$) & 
$(\mathbf{\bar{4}},\mathbf{1},\mathbf{\bar{2}})$ & (9,1,9) \\
($(\mathbf{\bar{4}},\mathbf{1},\mathbf{\bar{2}})_F$, 
$(\mathbf{15},\mathbf{\bar{2}},\mathbf{2})_H$) & 
($(\mathbf{4},\mathbf{2},\mathbf{1})_F$, 
$(\mathbf{15},\mathbf{1},\mathbf{3})_H$) & 
$(\mathbf{4},\mathbf{2},\mathbf{1})$ & (9,1,9) \\
\hline
\end{tabular}
\end{center}
\caption{
Dimension-five operators within PS unification (embedded in $SO(10)$ GUTs) and resulting predictions for the GUT scale ratios $y_e/y_d$, $y_u/y_d$ and $y_\nu/y_u$, where $\nu$,
$e$, $d$ and $u$ stand for any neutrino, charged lepton, down-type and up-type 
quark of the same generation.
$A, B, C, D$ and $X$ correspond to the fields in the supergraph for Yukawa couplings in figure \ref{fig:Messenger} which generates the dimension-five operator after integrating out the heavy messenger field. 
If the messenger representation $X$ has an index, there is more than one way to combine the fields $A$ and $B$ or $C$ and $D$ to form this representation leading to different predicted ratios. 
\label{tab:PSmessenger}}
\end{table}

\begin{table}
\begin{center}
\begin{tabular}{|c|c|}\hline
Operator dimension &	($y_e/y_d$, $y_u/y_d$) \\ \hline
\hline 4	&	(1,1)     \\
	&	(-3,1)     \\
\hline 5	&	(1,1)     \\
	&	(-3,1)     \\
	&	(9,1)     \\
\hline 6	&	(0,1/2)   \\
	&	(0,$\pm$1)     \\
	&	(0,2)     \\
	&	(3/4,0)   \\
	&	(3/4,1/2) \\
	&	(3/4,$\pm$1)   \\
	&	(3/4,2)   \\
	&	(1,0)     \\
	&	(1,1/2)   \\
	&	(1,$\pm$1)     \\
	&	(1,2)     \\
	&	(2,0)     \\
	&	(2,1/2)   \\
	&	(2,$\pm$1)     \\
	&	(2,2)     \\
	&	(-3,0)     \\
	&	(-3,1/2)   \\
	&	(-3,$\pm$1)     \\
	&	(-3,2)     \\

\hline
\end{tabular}
\end{center}
\caption{Summary of possible predictions from PS unification (embedded in $SO(10)$ GUTs) for the GUT scale ratios $y_e/y_d$ and $y_u/y_d$, where 
$e$, $d$ and $u$ stand for any charged lepton, down-type and up-type 
quark of the same generation. 
The predictions from certain dimension-six operators (taken from \cite{Allanach:1996hz}) are also included. 
 \label{tab:PSrelations}}
\end{table}

We now turn to the case of classes of $SO(10)$ GUTs where the breaking to the SM proceeds at $M_{GUT}$ via the PS breaking chain. 
At the stage of PS unified theories, the fermions of the SM are embedded in representations  $(\mathbf{4},\mathbf{2},\mathbf{1}) $ and $(\mathbf{\bar{4}},\mathbf{1},\mathbf{\bar{2}})$ of the PS gauge group as
\begin{eqnarray}
  F^{i \alpha a} & = & (\mathbf{4},\mathbf{2},\mathbf{1})^i =
         \begin{pmatrix}
         u_L^R & u_L^B & u_L^G & \nu_L \\
         d_L^R & d_L^B & d_L^G & e_L^-
         \end{pmatrix}^i , \\
  \bar{F}^{i}_{\alpha x} & = &
         (\mathbf{\bar{4}},\mathbf{1},\mathbf{\bar{2}})^i
       = \begin{pmatrix}
         \bar{d}_R^R & \bar{d}_R^B & \bar{d}_R^G & e_R^+ \\
         \bar{u}_R^R & \bar{u}_R^B & \bar{u}_R^G & \bar{\nu}_R
         \end{pmatrix}^i ,
\end{eqnarray}
where $\alpha = 1,\ldots,4$ is an $SU(4)_C$ index, $a,x = 1,2$ are
$SU(2)_{L,R}$ indices and $i=1,2,3$ is a family index. The fields in
$F^i$ form $SU(2)_L$ doublets and the fields in $\bar{F}^i$ 
$SU(2)_L$ singlets as indicated by the index L and R.
The MSSM Higgs $SU(2)_L$ doublets $h_u$ and $h_d$ are contained in
the bi-doublet representation
\begin{eqnarray}
  h_a^x &=& (\mathbf{1},\bar{\mathbf{2}},\mathbf{2}) = \begin{pmatrix}
         h_u^+ & h_d^0 \\
         h_u^0 & h_d^-
         \end{pmatrix}.
\end{eqnarray}
It acquires the vevs $v_u$ and $v_d$ in the $h_u^0$ and $h_d^0$ directions, respectively, which break the electroweak symmetry.
The breaking of the PS gauge symmetry to the SM can be achieved with the Higgs representations
\begin{eqnarray}
  H^{\alpha b} &=& (\mathbf{4},\mathbf{1},\mathbf{2}) = \begin{pmatrix}
         u_H^R & u_H^B & u_H^G & \nu_H \\
         d_H^R & d_H^B & d_H^G & e_H^-
         \end{pmatrix}, \\
  \bar{H}_{\alpha x} &=& (\bar{\mathbf{4}},\mathbf{1},\bar{\mathbf{2}}) = \begin{pmatrix}
         \bar{d}_H^R & \bar{d}_H^B & \bar{d}_H^G & e_H^+ \\
         \bar{u}_H^R & \bar{u}_H^B & \bar{u}_H^G & \bar{\nu}_H
         \end{pmatrix},
\end{eqnarray}
obtaining GUT scale vevs $\langle {\nu}_H \rangle$ and $\langle {\bar{\nu}}_H \rangle$.

Alternative to the bi-doublet and the quartets, other representations can contain the MSSM Higgs fields or can break the PS group to the SM. For example, the PS representation 
$(\mathbf{15},\bar{\mathbf{2}},\mathbf{2})$ can contain Higgs $SU(2)_L$ doublets which can develop an electroweak scale vev. This representation leads to the GJ relation in PS. 
Regarding the predictions for the neutrino Yukawa couplings, dimension
four operators with $(\mathbf{1},\bar{\mathbf{2}},\mathbf{2})_H$ ($(\mathbf{15},\bar{\mathbf{2}},\mathbf{2})_H$)
lead to the relation $y_\nu/y_u = 1$ ($y_\nu/y_u = -3$). 

Furthermore, the PS Higgs representations $(\mathbf{1},\mathbf{1},\mathbf{3})$, $(\mathbf{15},\mathbf{1},
\mathbf{1})$ and $(\mathbf{15},\mathbf{1},\mathbf{3})$ can arise from the common $SO(10)$ representations and have singlet components which can develop a GUT scale vev. Their inclusion in the effective operators which generate the Yukawa couplings can lead to new relations for the GUT scale Yukawa coupling ratios. 
We note that there are other fields like $SU(4)_C$ sextets or complete singlets which we do not consider here explicitly since they do not lead to new predictions. 

In table \ref{tab:PSmessenger} we have listed the possible
combinations of external and messenger fields which can appear
in the supergraph diagram of figure \ref{fig:Messenger}. 
The results for the GUT scale Yukawa ratios $y_e/y_d$ and $y_u/y_d$, where $e$, $d$ and $u$
stand for any charged lepton, down-type and up-type quark of the same
generation, are presented in table \ref{tab:PSrelations}. 
Furthermore, we also list the results
for certain dimension-six operators from Ref.~\cite{Allanach:1996hz}, which contain only the fields 
$F$, $\bar{F}$, $h$, $H$ and $\bar{H}$.

\newpage

\section{Phenomenological constraints on GUT scale mass ratios}
In this second part of the paper we analyse which ratios between quark and lepton masses (or Yukawa couplings) can be realised at the GUT scale when phenomenological constraints are taken into account. 
For explicitness, we will consider three minimal, but characteristic SUSY breaking scenarios, namely mAMSB, mGMSB and CMSSM, which provide boundary conditions for the soft SUSY parameters as we will briefly review in section \ref{sec:SUSYbreaking}. After RG evolution to low energies and including SUSY threshold corrections (see section \ref{sec:technical}), we apply the phenomenological constraints which we will describe in section \ref{sec:expcon}.
We note that we have not included neutrino masses in our analysis since we focus on Yukawa coupling ratios for charged fermions and since right-handed (s)neutrinos are also not included in the minimal SUSY breaking scenarios mAMSB, mGMSB and CMSSM. 

The GUT scale values of the quark and lepton masses, as well as of their ratios, can depend strongly on the SUSY threshold corrections. Particularly important in the large (or intermediate) $\tan \beta$ regime of the MSSM are the $\tan \beta$-enhanced threshold corrections for the down-type quarks and charged leptons.
In our analysis, we carefully include them for all families and types of charged fermions. 
The SUSY threshold corrections, in turn, depend on the SUSY parameters which are predicted from the considered SUSY breaking scenarios and which are subject to the phenomenological constraints.

Performing the above-described analysis, we arrive at phenomenologically allowed GUT scale ratios within the considered parameter ranges of the SUSY breaking scenarios mAMSB, mGMSB and CMSSM. These results are independent of any underlying GUT model. Finally, in section 4, we will compare them with the GUT predictions considered in the first part of the paper.

\subsection{Framework: Minimal SUSY breaking scenarios} \label{sec:SUSYbreaking}

SUSY, if realised in nature, obviously has to be broken in order to be
consistent with the experimental non-observation of sparticles so far. 
To keep SUSY as a solution to the hierarchy problem this breaking should be soft. 

In our analysis we will consider three common and characteristic examples for 
supersymmetry breaking scenarios, namely mAMSB \cite{AMSB}, 
mGMSB \cite{Ambrosanio:1997rv} and CMSSM \cite{mSUGRA, mSUGRArev} which 
provide boundary conditions for the soft SUSY breaking parameters at high energies. 
We will in all schemes choose the sign of $\mu$ to be positive in order to improve consistency  
with the experimental results on $(g-2)_\mu$, which we will discuss in section \ref{sec:g2mu}. 
The absolute value of $\mu$ is determined numerically to achieve successful electroweak symmetry breaking.

\subsubsection{mAMSB}

In the proposal for anomaly mediated SUSY breaking (AMSB), 
SUSY is broken on a separate brane and then mediated to the visible world
via the superconformal anomaly \cite{AMSB}. 
The parameter $m_{3/2}$, the vev of the auxiliary field in the
supergravity multiplet, determines the overall mass scale of the SUSY
particle masses. However, in the simplest AMSB model the sleptons are
tachyonic.
To cure this problem, in the minimal AMSB scenario (mAMSB) an additional
universal scalar soft mass $m_0$ is introduced.
The spectrum is then completely
determined by three parameters $m_{3/2}$, $m_0$, $\tan \beta$ and the
sign of $\mu$. 

Explicitly, the boundary conditions at the GUT scale in mAMSB
are given by 
\begin{eqnarray}
 M_a (M_{{GUT}}) & = & \frac{\beta(g_a)}{g_a} m_{3/2} \\
 A_y (M_{{GUT}}) & = & -\frac{\beta_y}{y}  m_{3/2} \\
 \tilde{m}_{\tilde{f}}^2 (M_{{GUT}}) & = & - \frac{1}{4} \left[
                \beta(g_a) \frac{\partial \gamma_{\tilde{f}}}{\partial g_a}
                + \beta_y \frac{\partial \gamma_{\tilde{f}}}{\partial y}
                \right] m_{3/2}^2 + m_0^2
\end{eqnarray}
where $a=1,2,3$, $M_a$ are the gaugino masses, $A_y$ the
trilinear couplings, $\tilde{m}_{\tilde f}$ the sfermion soft mass parameters, $\beta(g_a)$
is the $\beta$ function of the corresponding gauge coupling $g_a$,
$\beta_y$ the $\beta$ function of the Yukawa coupling $y$ and
the $\gamma_{\tilde{f}}$ is the anomalous dimension of the superfield
$\tilde{f}$. The mAMSB parameter ranges we will use in our analysis are given 
in table \ref{tab:mAMSB}.

\begin{table}
\begin{center}
\begin{tabular}{|c|c|c|c|}\hline
Parameter 		& Minimum & Maximum & Stepwidth \\ \hline
\hline $m_0$ in TeV 	& 0   & 3   & 0.1 \\ 
\hline $m_{3/2}$ in TeV	& 20  & 200 & 10  \\ 
\hline $\tan \beta$ 	& 20  & 60  & 2.5  \\ 
\hline
\end{tabular}
\end{center}
\caption{Parameter ranges and stepwidth used in our numerical scan
         for the mAMSB scenario. \label{tab:mAMSB}}
\end{table}

\subsubsection{mGMSB}

In the so-called minimal gauge mediated SUSY breaking scenario (mGMSB)
\cite{Ambrosanio:1997rv} the SUSY spectrum depends on four parameters:
the messenger mass $m_{\mathrm{mess}}$, the number of $\mathbf{5} \oplus \bar{\mathbf{5}}$ messenger fields $n_5$, the soft SUSY breaking mass scale $\Lambda$, the constant $c_{\mathrm{grav}}$ needed to calculate the
gravitino mass, $\tan \beta$ and the sign of $\mu$. We can set $c_{\mathrm{grav}} = 1$ without loss of generality, since we will not investigate observables depending on the gravitino mass. 
Since SUSY breaking is mediated via gauge
interactions, the soft scalar masses are predicted universal at $\Lambda$.

The universal boundary conditions are applied at the messenger scale
for the gaugino masses $M_a$, $a=1,2,3$ and the sfermion soft mass parameters 
$\tilde{m}_{\tilde f}$
\begin{eqnarray}
 M_a (m_{\mathrm{mess}}) & = & \frac{g^2_a}{16 \pi^2} n_5 \Lambda
        \tilde{g} \left( \frac{\Lambda}{m_{\mathrm{mess}}} \right), \\
 \tilde{m}_{\tilde f} (m_{\mathrm{mess}}) & = & 2 \Lambda^2 \sum_a 
        \left( \frac{g_a^2}{16 \pi^2} \right)^2 C_a n_5 
        \tilde{f} \left( \frac{\Lambda}{m_{\mathrm{mess}}} \right),
\end{eqnarray}
where
\begin{eqnarray}
 \tilde{g}(x) & = & \frac{1}{x^2} \left[ (1+x)\ln(1+x) +
                (1-x) \ln (1-x) \right] \; ,\\
 \tilde{f}(x) & = & \frac{1+x}{x^2} \left[ \ln(1+x)
                -2 \mathrm{Li}_2 \left(\frac{x}{1+x}\right)
                +\frac{1}{2} \mathrm{Li}_2\left(\frac{2x}{1+x}\right)
                \right] + (x \rightarrow - x) \; ,
\end{eqnarray}
and where $C_a$ is the quadratic Casimir invariant of the MSSM scalar field
in question. The masses are run from $M_{GUT}$ to $M_Z$ via two-loop RGEs. 
We note that the running between the messenger scale and the GUT 
scale is performed using MSSM RGEs, which provides a good approximation.
The mGMSB parameter ranges we will use in our analysis are given 
in table \ref{tab:mGMSB}.

\begin{table}
\begin{center}
\begin{tabular}{|c|c|c|c|}\hline
Parameter 			& Minimum & Maximum & Stepwidth \\
\hline
\hline $n_5$		& 1   & 5   & 1 \\
\hline $\Lambda$ in TeV		& 10  & 200 & 20  \\
\hline $m_{\mathrm{mess}}$
	& $1.01 \Lambda$ & $10^5 \Lambda$ & $10^4 \Lambda$ \\
\hline $c_{\mathrm{grav}}$	& 1  & 1 & - \\
\hline $\tan \beta$		& 20  & 60  & 2 \\
\hline
\end{tabular}
\end{center}
\caption{Parameter ranges and stepwidth used in our numerical scan
         for the mGMSB scenario. \label{tab:mGMSB}}
\end{table}

\subsubsection{CMSSM}

In the constrained MSSM (CMSSM) SUSY breaking scenario \cite{mSUGRA, mSUGRArev} 
the soft SUSY breaking parameters are assumed to be universal at the GUT scale 
and therefore the SUSY particle spectrum is determined by four parameters and one sign: 
the scalar mass $m_0$, the fermionic mass $m_{1/2}$, the trilinear
coupling $A_0$, the ratio of the Higgs vevs $\tan \beta$ and the sign of $\mu$. 
The boundary conditions for the soft SUSY breaking parameters, imposed at the GUT scale, are
\begin{eqnarray}
 M_a (M_{{GUT}}) & = & m_{1/2} \\
 A_y (M_{{GUT}}) & = & A_0 \\
 \tilde{m}_{\tilde{f}}^2 (M_{{GUT}}) & = & m_0^2 \;,
\end{eqnarray}
where again $a=1,2,3$, $M_a$ are the gaugino masses,
$A_y$ the trilinear couplings and $\tilde{m}_{\tilde{f}}$ the
sfermion masses. The parameter range 
we will use in our analysis is given in table \ref{tab:CMSSM}. 

\begin{table}
\begin{center}
\begin{tabular}{|c|c|c|c|}\hline
Parameter 		& Minimum & Maximum & Stepwidth \\ \hline
\hline $m_0$ in TeV	& 0  & 3 & 0.2  \\
\hline $m_{1/2}$ in TeV	& 0  & 3 & 0.2 \\
\hline $A_0$ in TeV	& -3  & 3 & 1.5 \\
\hline $\tan \beta$	& 20  & 60 & 5  \\
\hline
\end{tabular}
\end{center}
\caption{Parameter ranges and stepwidth used in our numerical scan
         for the CMSSM scenario. \label{tab:CMSSM}}
\end{table}

\subsection{Numerical procedure and the role of SUSY threshold 
corrections} \label{sec:technical}
Using the soft breaking parameters specified in section \ref{sec:SUSYbreaking} as high scale boundary conditions, the MSSM parameters are run to low energies using a modified version of SoftSUSY 2.0.18 \cite{SoftSUSY} which we have also used for calculating the spectrum. 
SoftSUSY runs in loops to achieve consistency with high scale boundary conditions as well as with low scale input, thereby determining $|\mu|$. From SoftSUSY we read out the masses of the quarks and charged leptons at the GUT scale. Our modification to the SoftSUSY code are the following:
\begin{itemize}

\item In SoftSUSY 2.0.18, the threshold corrections are included 
as self-energy corrections to the fermion masses, but only for the
third family. We have included the SUSY threshold corrections 
for the first two generations, using mainly the formulae of \cite{BPMZ}.
The large logs appearing in the formulae in \cite{BPMZ}
are already resummed in the gauge couplings and therefore are not
included anymore (see also \cite{SoftSUSY}). For the first two
generations we have also set the external momenta of the fermions to
zero. 
This provides a very good approximation since corrections are of the 
order of $m_f/M_{SUSY}$, where $m_f$ is the mass
of the corresponding (light) fermion and $M_{SUSY}$ is the mass scale
of the SUSY particles involved in the loops. We have also updated the 
experimental data on the quark masses according to \cite{Xing:2007fb}. 

\item We have furthermore modified SoftSUSY 2.0.18 to include left-right mixing
for the first two families, which was set to zero. 
The left-right mixing angle $\theta_{\tilde{f}}$ is defined (at tree-level)
as
\begin{equation} \label{eq:mixingangle}
\sin ( 2 \theta_{\tilde{f}} ) = \frac{2 m_f (A_f - \mu \tan \beta)}
                                {m_{\tilde{f}_1}^2 - m_{\tilde{f}_2}^2},
\end{equation}
where $f=e,\mu,\tau,d,s,b$. $A_f$ is the corresponding trilinear
coupling and $m_{\tilde{f}_{1/2}}^2$ are 
the corresponding mass eigenvalues of the sfermion mass matrix. 
For our study it was necessary to include it since we found that 
for some parameter points it is not negligible.  
For example, in the mAMSB scenario for $m_0 = 500$~GeV, $m_{3/2} = 20$~TeV 
and $\tan \beta = 30$ we obtain
$\theta_{\tilde{s}}  \approx 0.58$ and
$\theta_{\tilde{b}} \approx 0.35$. 
This large mixing can be understood from the fact that the splitting
between the sfermion mass eigenstates in the example is mainly driven 
by the mass of the fermion. Then both, the numerator and the denominator
of Eq.~(\ref{eq:mixingangle}) are small, leading to sizable mixing. 
 
\item Some of the points in our parameter scan are already excluded
by SoftSUSY and are not displayed in our results. 
This happens for example if the spectrum contains tachyons or if it is
not possible to achieve a successful electroweak
symmetry breaking (see SoftSUSY manual \cite{SoftSUSY}). 
In addition, we have also made SoftSUSY reject parameter points where
the calculated SUSY threshold corrections are so large that the
perturbative expansion is spoiled. 
\end{itemize}  

Regarding the calculation of the experimental constraints, for some of 
them we have exported the spectrum calculated from SoftSUSY to micrOMEGAs 2.2 CPC
\cite{micromegas} using the SLHA \cite{SLHA} interface. 
The experimental constraints we will use in our analysis are discussed in 
detail in the next section.

\subsection{Experimental constraints} \label{sec:expcon}

\subsubsection{Direct detection}

The LEP experiments have searched for SUSY
particles with negative results \cite{PDG}. In our analysis we exclude
parameter points with a chargino or slepton (sneutrino and charged
slepton) lighter than the LEP bounds.
We have not applied the LEP bound for the Higgs boson mass which holds only
in SM (or approximately for a SM-like Higgs). 
However, for almost all parameter points which pass the remaining constraints we 
have checked that the lightest CP-even Higgs boson was heavier than the LEP bound
and for the other parameter points it was still above 105 GeV. 
For these points there may be some tension with the LEP data. However, for the
outcome of our study it makes no difference if they are included or excluded.

\subsubsection{Electroweak precision observables}

We have furthermore included constraints from electroweak precision observables (EWPO) such as the W boson
mass $M_W$ and the effective leptonic weak mixing angle
$\sin^2 \theta_{\mathrm{eff}}$. These observables are known to a high
accuracy from LEP and Tevatron.

In \cite{MWexp} a combined world result for the W boson
mass of
\begin{equation}
 M_W = 80.429 \pm 0.039 \; \mathrm{GeV}
\end{equation}
is given 
and in \cite{sintheffexp} the up-to-date experimental result for
the effective leptonic weak mixing angle is listed as
\begin{equation}
 \sin^2 \theta_{\mathrm{eff}} = 0.23153 \pm 0.00016 \;.
\end{equation}
By applying these results as a constraint we demand that the theoretical
predictions for a given parameter point (calculated by SoftSUSY) 
lie within the above given $1\sigma$ errors.

\subsubsection{$\boldsymbol{\mathrm{BR}(b \rightarrow s \gamma)}$}

The decay $b \rightarrow s \gamma$ occurs in the SM as well as in the
MSSM at one loop level, which makes it very
interesting as a probe of physics beyond the SM.
The present experimental value for
$\mathrm{BR}(b \rightarrow s \gamma)$, released by the Heavy Flavour
Averaging Group (HFAG), is \cite{bsgexp}
\begin{equation}
 \mathrm{BR}(b \rightarrow s \gamma) 
   = \left( 3.55 \pm 0.24^{+0.09}_{-0.10} \pm 0.03 \right)
     \times 10^{-4} \; ,
\end{equation}
where the first error is the combined statistical and uncorrelated
systematic uncertainty, and the other two errors are correlated
systematic theoretical uncertainties and corrections respectively.

We evaluate $\mathrm{BR}(b \rightarrow s \gamma)$ for our data points
using micrOMEGAs \cite{micromegas} and exclude the data points which do not lie
within the interval $\left( 3.55^{+0.36}_{-0.37} \right) \times 10^{-4}$.
For our analysis we use the summed errors to define the allowed region.

\subsubsection{$\boldsymbol{\mathrm{BR}(B_s \rightarrow \mu^+ \mu^-)}$}

The present experimental upper limit on
$\mathrm{BR}(B_s \rightarrow \mu^+ \mu^-)$ from the Fermilab Tevatron
collider is $5.8 \times 10^{-8}$ at the 95 \% C.L. \cite{bsmmexp}.
The SM prediction for this branching ratio is
$\left( 3.4 \pm 0.5 \right) \times 10^{-9}$ \cite{bsmmtheosm}, leaving 
some room for a possible large SUSY contribution. We have calculated this
contribution using the micrOMEGAs package. We impose the constraint that
the SUSY contribution does not exceed the experimental bound minus
the lower limit of the SM contributions.

An approximate formula for the
SUSY corrections to $\mathrm{BR}(B_s \rightarrow \mu^+ \mu^-)$ is \cite{bsmmtheosusy}
\begin{eqnarray}
\mathrm{BR}(B_s \rightarrow \mu^+ \mu^-) & \simeq & 3.5 \times 10^{-5}
\left[ \frac{\tan \beta}{50} \right]^6 \left[ \frac{\tau_{B_s}}{1.5 \mathrm{ps}} \right]
 \left[ \frac{F_{B_s}}{230 \mathrm{MeV}} \right]^2 \left[ \frac{|V_{ts}|}{0.04} \right]^2
  \nonumber\\
  & \times & \frac{\overline{m}_t^4}{M_A^4}
   \frac{(16 \pi^2 \epsilon_Y)}{(1+ \tilde{\epsilon}_3 \tan \beta)^2(1+ \epsilon_0 \tan \beta)^2}
\end{eqnarray}
where $\overline{m}_t \equiv m_t(\mu_t)$ and
$\tilde{\epsilon}_3 = \epsilon_0 + y_t^2 \epsilon_Y$. The full
expressions for $\epsilon_0$ and $\epsilon_Y$ can be found
in \cite{bsmmtheosusy}. The branching ratio is proportional to $\tan^6 \beta$
as well as to
$\epsilon_Y$, which in turn is proportional to the trilinear coupling of the stops.
This means that large $\tan \beta$ and a large trilinear coupling pushes the
branching ratio to larger values whereas a heavier CP-odd Higgs
boson can suppress the branching ratio.

\subsubsection{Dark matter} \label{sec:CDM}

In the MSSM (with R-parity conserved) the lightest supersymmetric particle (LSP) provides an interesting candidate for the dark matter particle. It may be the lightest neutralino, but may alternatively be the gravitino. 
The WMAP Collaboration, after five years of data taking, 
has released $\Omega_m h^2 = 0.1143 \pm 0.0034$ for the dark matter 
density in the Universe \cite{WMAP5}.

If one makes the assumption of a ``standard'' cosmological evolution as well as 
that dark matter dominantly consists of the lightest neutralino, this would 
imply rather strong constraints on the parameter space of SUSY models. However, 
other particles may contribute to dark matter in addition to a
neutralino LSP, which relaxes this bound to the requirement that the relic density 
of the neutralino, which we require to be the LSP, should not exceed the dark matter observed by WMAP.  

We will discuss this relaxed bound separately in the following, since it may be 
taken as a possible constraint under additional assumptions. However, since it can be 
avoided if, for instance, the cosmological evolution is ``nonstandard'' or if a small amount of R-parity violation is introduced, we do not include it in our final results. 
Furthermore, in mGMSB the gravitino is the LSP and its relic density depends on its mass, which we treat as a free parameter in this setup such that no constraint can be applied.

\subsubsection{Anomalous magnetic moment of the muon} \label{sec:g2mu}

The results for the anomalous magnetic moment of the muon
$(g-2)_\mu$ (or for the parameter $a_\mu = \frac{1}{2}(g-2)_\mu$, respectively) are still not completely
settled. In particular there is some tension between the preliminary $\tau$ data from
BELLE \cite{Hayashii:2005ih} and the $e^+ e^-$ data
\cite{Davier:2007ua} for the hadronic contributions (for a review see
e.g.\ \cite{gmu}). With the $e^+ e^-$ data for the hadronic
contributions and the final result of the Brookhaven E821 experiment
\cite{gmuexp} the difference between the experiment and the
theoretical SM prediction is
\begin{equation}
 a_\mu^{\mathrm{exp}} - a_\mu^{\mathrm{theo}}
   = (27.5 \pm 8.4) \times 10^{-10}
\end{equation}
equivalent to a 3.3$\sigma$ deviation. Three other recent evaluations
yield slightly different numbers \cite{gmualt}.
Because of the discrepancies between the electron and the $\tau$ data
and the slight differences in the theoretical predictions we only
use as constraint that the SUSY contributions to $(g-2)_\mu$
have the right sign to make
$a_\mu^{\mathrm{exp}} - a_\mu^{\mathrm{theo}}$ smaller and that they are 
not too large, $0 \leq a_\mu \leq 35.9 \times 10^{-10}$.

For the calculation of $(g-2)_\mu$ we use micrOMEGAs which has implemented 
the formulae from
\cite{gmuformulas}. There is also an approximate formula given in
\cite{gmu} for the case that all SUSY parameters are set to
$M_{SUSY}$, $\mathrm{sgn} (M_1) = \mathrm{sgn} (M_2)$ and all
parameters are real:
\begin{equation}
\delta a_\mu^{SUSY} \approx 13 \tan \beta \:\mathrm{sgn} (\mu M_{1,2})
  \left( \frac{100 \mathrm{GeV}}{M_{SUSY}} \right)^2 10^{-10}.
\end{equation}
From this formula we already see, that large values of $\tan \beta$
can lead to conflicts with experimental observations, if also the SUSY
scale is not too large. The anomalous magnetic moment receives also
larger corrections for smaller smuon and muon-sneutrino masses and
larger neutralino and chargino masses. Furthermore, we can also see the
dependence on the sign of $\mu$. For example, our constraints exclude 
a negative $\mu$ if both $M_1$ and $M_2$ are positive.

\subsection{Allowed quark and lepton mass ratios at the GUT scale}\label{sec:results_constraints}\label{sec:Results}\label{sec:final}

Performing the numerical scan over the parameter ranges for the SUSY breaking scenarios specified in \ref{sec:SUSYbreaking}, we obtain the scatter plots with allowed GUT scale values for the quark and lepton mass ratios of interest shown in figures 2 - 5. 
For each of the parameter points, corresponding to specific boundary conditions for the SUSY breaking parameters at high energies, we apply the experimental constraints from direct searches, EWPO,
$BR(B_s \rightarrow \mu^+ \mu^-)$, $BR(b \rightarrow s \gamma)$ and
$(g-2)_\mu$ described in section \ref{sec:expcon}. 
Values shown in black are consistent with the applied constraints, whereas dots in red mark parameter points which are excluded. 
The grey regions around the black dots indicate the allowed ratios when the experimental (1$\sigma$) errors on the quark masses are included.  The other lines and dots correspond to possible GUT predictions and will be discussed in section \ref{sec:theoryvsexp}. We now discuss the impact of the experimental constraints in the considered SUSY breaking scenarios.

\subsubsection{mAMSB}\label{sec:constraints_mAMSB}

The first row in figure \ref{fig:finalresults} shows the combined results for mAMSB. For the considered mAMSB parameter range (see section \ref{sec:SUSYbreaking}) we can see from the left plot that, with quark mass errors included, $m_\mu / m_s$ in the range from 2.48 to 5.72 and  $m_e / m_d$  in the range from 0.21 to 0.65 are possible. The right plot shows that for $y_\tau / y_b$ values in the range from 0.98 to 1.3 and for $y_t / y_b$ in the range from 1.37 to 4.78 are allowed. Compared to the yellow squads indicating the values calculated  without taking the SUSY threshold corrections into account, we see that all ratios are reduced. As discussed in \cite{Ross:2007az,Antusch:2008tf}, the reason for this is that the sign of the dominant $\tan \beta$-enhanced correction parameter $\varepsilon_i^G$ is negative for negative gluino mass $M_3$ when $\mu$ is positive, which enhances the down-type Yukawa couplings at the SUSY scale and finally lowers the possible values of the ratios at $M_{GUT}$.  
Large SUSY threshold corrections, and thus lower values of the GUT scale ratios correspond to large $\tan \beta$. 
The plots also show that there is a strong correlation between $m_\mu / m_s$ and $m_e / m_d$, which stems from the fact that the masses of the first two sfermion generations are very similar.

One can see from the plots how the phenomenological constraints restrict the possible effects of the threshold corrections on the GUT scale ratios. 
First of all, a sparticle spectrum free of tachyons already excludes values of $m_0$ below about 200 GeV. 
Furthermore, we found that large values of $\tan \beta$ above 50 did not lead to a viable spectrum. 
 These parameter points were rejected by the numerics and are not displayed in figure \ref{fig:finalresults}. 
In the parameter range we considered (and for points with a tachyon-free spectrum), the strongest constraint was $b \rightarrow s \gamma$ (c.f.\ the first row of figure \ref{fig:plots_mAMSB}), which disfavours large values of $\tan \beta$. In mAMSB, EWPO (c.f.\ the second row of figure \ref{fig:plots_mAMSB}) also provide a significant constraint and disfavour large values of $\tan \beta$. Compared to $b \rightarrow s \gamma$ and EWPO, the limits from $B_s \rightarrow \mu^+ \mu^-$ and $(g-2)_\mu$ are much less constraining. Including all constraints the minimal allowed $m_0$ raises to about 600~GeV and the maximal $\tan \beta$ reduces to about 45.

Finally, under the assumption that the neutralino is the LSP, stable due to R-parity, and that the evolution of the universe is standard up to temperatures where the LSP freezes out, the LSP relic density could be used as an additional constraint. In particular the parameter points which lead to a LSP relic density larger than the dark matter density or where the LSP is charged would be excluded. The impact which this constraint would have is shown in the third row of figure \ref{fig:plots_mAMSB}. The consequence would be that only a small region where the threshold corrections are comparatively small would remain allowed.

\subsubsection{mGMSB}

The combined results for mGMSB are shown in the second row in figure \ref{fig:finalresults}. Compared to the case of mAMSB and following the arguments of section \ref{sec:constraints_mAMSB}, positive $M_3$ with positive $\mu$ leads to a positive threshold correction parameter $\varepsilon_i^G$ which lowers the down-type Yukawa couplings and consequently enlarges the Yukawa coupling ratios compared to the case without threshold effects included. For the considered mGMSB parameter range (see section \ref{sec:SUSYbreaking}) we can see from the left plot that, with quark mass errors included, $m_\mu / m_s$ in the range from 3.62 to 7.69 and  $m_e / m_d$  in the range from 0.30 to 0.87 are possible. The right plot shows that for $y_\tau / y_b$ values in the range from 1.35 to 2.09 and for $y_t / y_b$ in the range from 1.01 to 5.26 are allowed.

Turning to the individual experimental constraints, in mGMSB with the parameter range specified in section \ref{sec:SUSYbreaking} all applied constraints lead to a significant reduction of the possible GUT scale ratios. As in mAMSB, the strongest constraint is $b \rightarrow s \gamma$ (c.f.\ the first row of figure \ref{fig:plots_mGMSB}), followed by EWPO (c.f.\ the second row of figure \ref{fig:plots_mGMSB}) and $(g-2)_\mu$ and finally by limits from direct searches and $B_s \rightarrow \mu^+ \mu^-$.  
We note that due to the correlation between $m_\mu / m_s$ and $m_e / m_d$ many parameter points lead to the same ratio which means that the dots would lie on top of each other. If at least one of the parameter points is consistent with the phenomenological constraints, the ratio is shown in black. 
 
Dark matter constraints are not discussed since the gravitino is generically the LSP in GMSB and the gravitino mass essentially represents a free parameter in our setup.

\subsubsection{CMSSM}

In the CMSSM, as in mGMSB, with positive $M_3$ and $\mu$ the SUSY threshold corrections tend to reduce the down-type Yukawa couplings and consequently enlarge the Yukawa coupling ratios at the GUT scale. The combined results for CMSSM  are shown in the third row of figure \ref{fig:finalresults}. For the CMSSM parameter ranges specified in section \ref{sec:SUSYbreaking} we find that, with quark mass errors included, $m_\mu / m_s$ can be in the range from 3.44 to 7.73  and  $m_e / m_d$ in the range from 0.29 to 0.87. The right plot shows that for $y_\tau / y_b$ values in the range from 1.28 to 2.10 and for $y_t / y_b$ in the range from 0.97 to 5.71 are allowed.

The first and second row of figure \ref{fig:plots_CMSSM} show as examples the impact of the constraints from 
$b \rightarrow s \gamma$ and EWPO. The main consequence regarding the allowed GUT scale ratios is that points are excluded where the SUSY threshold corrections tend to reduce the GUT scale ratios. This is in agreement with \cite{Altmannshofer:2008vr}, where it has been argued that third family Yukawa coupling unification within the inverted scalar mass hierarchy scenario \cite{ISMH} requires a region of parameter space where $-A_0 \approx 2 m_0$ and $\mu, m_{1/2} \ll m_0$ and that this inevitably leads to conflicts with bounds on, e.g., $B_s \rightarrow \mu^+ \mu^-$ because of the large trilinear coupling. We note that we have not focused on this specific correlation between the parameters which explains why we have only relatively few (excluded) parameter points which are close to third family Yukawa unification.

The third row of figure \ref{fig:plots_CMSSM} shows the constraints which would come from the requirement that the neutralino relic density does not exceed the observed dark matter density, under the assumptions that the neutralino is the stable LSP and that the cosmic history is ``standard''. We find from our scan that the impact of this constraint would be that a certain region with large $\tan \beta$ would be favoured. However, we would like to note that there are comparatively thin parameter space regions which lead to a viable neutralino relic density, i.e.\ the so-called funnel and coannihilation regions. Since our parameter space is comparatively coarse, we cannot exclude that we have missed viable parameter points in these thin regions. Such points could lead to additional possibilities for allowed GUT scale ratios. The few points with larger $y_t/y_b$ (i.e.\ smaller $\tan \beta$) belong to these thin parameter space regions. The dark matter constraints, which only apply under additional assumptions, are not included in the final results.

\newpage

\section{Allowed GUT scale ratios compared to theory predictions}\label{sec:theoryvsexp}

As discussed in the previous section, within mAMSB, mGMSB and CMSSM only certain ranges of GUT scale ratios 
$m_e/m_d$, $m_\mu/m_s$, $y_\tau/y_b$ and $y_t/y_b$ are allowed when phenomenological constraints from electroweak precision observables, $B$ physics, $(g-2)_\mu$ and mass limits on sparticles are taken into account. In this section we compare these ranges with the possible predictions for these ratios from unified theories. Figure \ref{fig:finalresults} contains our final results. The red dots correspond to parameter points which are excluded by phenomenological constraints, while the black dots are allowed with grey regions indicating the experimental (1$\sigma$) errors on the quark masses.

The possible theory predictions discussed in section 2 are shown in figure \ref{fig:finalresults} as green and blue lines and dots. We note that for mass ratios only the modulus of the ratio is relevant, since a sign only corresponds to a global phase redefinition. We will therefore in the following always display the modulus of the predicted ratios. 
The different colours have the following meaning: Green lines denote the predictions from $SU(5)$ GUTs (eventually embedded in $SO(10)$) and dashed green and blue lines the predictions which can arise in PS unification (eventually embedded in $SO(10)$ GUTs) as well as in GUTs based on $SU(5)$. For the third family the dark blue points denote the predictions from operators up to dimension five in PS unification, whereas the light blue points denote predictions which can arise from certain dimension-six operators.

\subsection{GUT predictions vs phenomenological constraints in mAMSB}

From figure \ref{fig:finalresults} we see that mAMSB is the only 
considered scenario where the GJ relation $m_\mu/m_s = 3$
is allowed. 
Its realisation requires intermediate $\tan \beta$ (around 30) and a comparatively heavy sparticle spectrum corresponding to $m_0$ above about 1 TeV and $m_{3/2}$ above about 100 TeV. 
Interestingly, this parameter region would also be
compatible (with quark mass errors included) with the second GJ relation $m_e/m_d = 1/3$,
which arises in the presence of a texture zero in the (1,1)-elements of the 
Yukawa matrices and under the assumption that they are symmetric.   

In addition to the GJ relation, mAMSB is also compatible with
the ratio $m_\mu/m_s = 9/2$. This ratio arises in all scenarios whenever 
the SUSY threshold corrections are comparatively small, for instance if 
$\tan \beta$ is small such that there is no $\tan \beta$ enhancement. 
In figure \ref{fig:finalresults} the yellow squad shows the GUT scale ratios
which would result when the SUSY threshold corrections were ignored. In the
absence of SUSY threshold corrections a value close to $m_\mu/m_s = 9/2$
would result as well.

Regarding the third generation we find that third family Yukawa 
unification $y_t = y_b = y_\tau$ is not compatible with mAMSB. The parameter points which 
came close to this relation were all excluded because either the spectrum contained tachyons 
and/or because it was not possible to achieve successful electroweak symmetry breaking. 
Partial third family Yukawa unification $y_\tau/y_b = 1$ turned out to be possible.  
Interestingly, $y_\tau/y_b = 1$ is realised in combination with $y_t/y_b = 2$.
Both relations can emerge simultaneously from a dimension-six operator within PS 
unified theories. 

The GUT predictions $y_t = 2 y_b = 2 y_\tau$ and $m_\mu/m_s = 3$ can be realised for the same region of parameter space where $\tan \beta$ is intermediate and the sparticle spectrum is rather heavy.  
We would like to note that including the dark matter constraint would exclude this parameter space region (see third row in figure 3). However, for example, in variants of mAMSB where a small amount of R-parity violation is introduced or in ``nonstandard'' cosmology, this constraint might be avoided.

\subsection{GUT predictions vs phenomenological constraints in mGMSB and CMSSM}

The allowed GUT scale ranges 
within mGMSB and CMSSM differ significantly from the ranges in mAMSB. This is 
due to the fact that the sign of the generically dominant $\tan \beta$-enhanced SUSY QCD threshold correction
is governed by $\mbox{sgn}(\mu M_3)$ which is positive in mGMSB and CMSSM but negative in mAMSB. 
It has turned out that mGMSB and CMSSM are in fact compatible with the same theory predictions.
We will therefore discuss both scenarios together in this subsection.

For mGMSB and CMSSM the GJ relation $m_\mu/m_s = 3$ is disfavoured. 
For small $\tan \beta$, i.e.\ small threshold corrections, both scenarios (and also mAMSB) are 
compatible with $m_\mu/m_s = 9/2$. 
In addition, for large $\tan \beta$ (i.e.\ large SUSY threshold corrections), the theory prediction 
$m_\mu/m_s = 6$ can be compatible with phenomenological constraints. The GUT scale ratios $m_\mu/m_s = 9/2$ as well as 
$m_\mu/m_s = 6$ can be realised in $SU(5)$ GUTs, however, within our setup, not from the PS gauge group. 

Regarding the third generation we again find that third family Yukawa 
unification $y_t = y_b = y_\tau$ is incompatible. 
However, interesting alternative relations are compatible with data: 
One example is the GUT scale prediction $y_\tau/y_b = 3/2$ which arises in the context of $SU(5)$ GUTs.
It can be realised for moderate values of $\tan \beta$ (e.g.\ $\tan \beta \approx 25$ in CMSSM) while it would be disfavoured for large values of $\tan \beta$. We would like to remark that this region of parameter space is also consistent with the GUT prediction $m_\mu/m_s = 9/2$. 
For large $\tan \beta$ (i.e.\ large SUSY threshold corrections), on the other hand, the relations  
$y_\tau/y_b = 2$ and $y_t/y_b = 1$ are allowed. Interestingly, the relation $ 2 y_t = 2 y_b = y_\tau$ can also emerge as a prediction from dimension-six operators within PS unified theories. 
The parameter space where $2 y_t = 2 y_b = y_\tau$ is realised additionally allows to realise the GUT relation $m_\mu/m_s = 6$. However, while $m_\mu/m_s = 6$ appears in $SU(5)$ the relation $2 y_t = 2 y_b = y_\tau$ can emerge from PS. 
In our scan we found no parameter point in mGMSB and CMSSM where partial third family Yukawa unification $y_\tau/y_b = 1$ was compatible with experimental constraints.

\subsection{Comparison with previous studies}

The viability of third family Yukawa unification $y_t = y_b = y_\tau$ (and also on the less restrictive possibility $y_b = y_\tau$) has been extensively studied in the literature (see e.g.~\cite{Carena:1994bv, Hempfling:1993kv,Ross:2007az,Antusch:2008tf,Altmannshofer:2008vr,Bagger:1996ei,King:2000vp, Blazek:2002ta}). The recent study \cite{Altmannshofer:2008vr} has reconsidered the phenomenological viability of this relation and it has been pointed out that in a variant of the CMSSM with nonuniversal soft Higgs mass parameters (NUHM) the relation $y_t = y_b = y_\tau$ is quite challenged by the experimental data from B physics. 
SUSY threshold effects on the GJ relations have been discussed recently in \cite{Ross:2007az,Antusch:2008tf}.

In \cite{Antusch:2008tf}, the impact of the $\tan \beta$-enhanced SUSY threshold corrections for all three generations and for down-type quarks as well as for charged leptons has been analysed numerically and analytically. For this purpose the threshold corrections have been treated in the EW-unbroken phase. The possible ranges for the GUT scale values of the Yukawa couplings and their ratios have been calculated for three example ranges of low energy SUSY parameters
and it has been pointed out that the presence of SUSY threshold corrections can open up new possibilities for GUT model building. 

Compared to \cite{Antusch:2008tf} our results are in good qualitative agreement (c.f.\ figure 5 of \cite{Antusch:2008tf} where the results are presented in a similar way). The example SUSY parameter range $a$ in \cite{Antusch:2008tf} was inspired by anomaly mediated SUSY breaking and the SUSY parameter ranges $g_+$ (and $g_-$) by scenarios with gaugino unification and $\mu >0$ ($\mu < 0$). 
Quantitatively there are nevertheless differences, which are larger for the third family. 
For example for $\tan \beta = 30$ in the mAMSB case we find that (before applying experimental constraints) $m_\mu / m_s$ can be in the range 2.41-5.73, whereas in \cite{Antusch:2008tf} in case $a$ we found the very similar range 2.40-5.63. 
On the other hand, for the ratio $y_\tau/y_b$ we find an allowed range
of 0.94-1.28 within mAMSB compared to 0.60-1.39 for the example SUSY parameter range $a$. 
However, since in the present study we are considering explicit SUSY breaking scenarios at high energy resulting in different low energy SUSY spectra, there is no reason to expect perfect quantitative agreement. 

The main difference from \cite{Antusch:2008tf} is of course that the consideration of explicit SUSY breaking scenarios allows to take phenomenological constraints into account. Their restrictions on the allowed GUT scale ratios depend somewhat on the explicit minimal SUSY breaking scenario, however we expect that some consequences are also characteristic for variants of the considered schemes. For example, it has turned out that there is a certain tension between realising GUT predictions which require large SUSY threshold corrections and the experimental constraints which basically restrict the effects of SUSY loops to the observables. It has also turned out that, contrary to claims in \cite{Ross:2007az,Antusch:2008tf}, it may be  challenging to realise third family Yukawa unification in AMSB-like SUSY breaking scenarios. 
Finally, we go beyond \cite{Antusch:2008tf} by investigating explicitly which alternative GUT scale predictions for quark and lepton mass ratios can emerge in unified theories and by comparing them to the phenomenologically allowed GUT scale ratios.

\subsection{Additional implications of our results}

\subsubsection{GUT scale ratios for the first fermion generation}
As mentioned in section 2, the relation between the mass of first generation of fermions and the Yukawa couplings is often more complicated. We have therefore focused on the second and third generation so far. 

As discussed in section 2, predictions for the ratios between quark and charged lepton masses at the GUT scale can arise if two conditions are satisfied: a hierarchical structure of the Yukawa matrices and the situation that one single GUT operator dominates the relevant Yukawa matrix element.   
The simplest case which can lead to predictions for the first generation of fermions is that the submatrix for the first and the second fermion generation is also hierarchical. Then the masses of the first fermion generation would be approximately determined by the diagonal elements (i.e.\ the (1,1)-elements) of the corresponding Yukawa matrices and the phenomenologically allowed range for $m_e/m_d$ can directly be compared to the theory predictions in tables 2 and 4 of section 2. The theory prediction $m_e/m_d = 1/2$, possible in $SU(5)$, or the relation $m_e/m_d = 3/4$ from PS unification would be compatible with the experimental constraints.

In many GUT models of fermion masses and mixings, however, a different situation is realised: There, the 
Yukawa matrices are symmetric and have a zero in the (1,1)-entries (see e.g.\ \cite{GST}). In this case, the mass of the electron and down-type quark are inversely proportional to the masses of the second generation and, in addition, depend on the (1,2)-entries (which are equal to the (2,1)-entries by assumption) of the Yukawa matrices. More precisely, the prediction for the ratio $m_e/m_d$ is then given by
\begin{eqnarray}
\frac{m_e}{m_d} = \frac{m_s}{m_\mu}\, \frac{(Y_e)^2_{12}}{(Y_d)^2_{12}}\;.
\end{eqnarray} 
For ${(Y_e)_{12}}/{(Y_d)_{12}}=1$ and $m_\mu/m_s = 3$ we recover the second 
GJ relation $m_e/m_d=1/3$ which is consistent with our results when quark mass errors are included. Interestingly, it is possible to realise both relations within mAMSB. 
With ${(Y_e)_{12}}/{(Y_d)_{12}}=1$, no alternative GUT prediction for $m_\mu/m_s$ is consistent with the above assumptions, due to the strong correlation between $m_e/m_d$ and $m_\mu/m_s$ in figure \ref{fig:finalresults}. 

However, with a different Clebsch factor relating $(Y_e)_{12}$ to $(Y_d)_{12}$, the alternative GUT predictions $m_\mu/m_s = 9/2$ and $m_\mu/m_s = 6$ can well be consistent with the assumption of symmetric Yukawa matrices with zero (1,1)-elements: 
The relation $m_\mu/m_s = 9/2$ is consistent with $m_e/m_d = 1/2$, which would require ${(Y_e)_{12}}/{(Y_d)_{12}}\approx 3/2$. 
Similarly, $m_\mu/m_s = 6$ is consistent with $m_e/m_d = 3/2$, which would require ${(Y_e)_{12}}/{(Y_d)_{12}}\approx 2$.  
Of course, when one of the above assumptions (i.e.\ symmetric Yukawa matrices and zero (1,1)-elements) is dropped then there are more possibilities. For example, without zero (1,1)-element  the relation ${(Y_e)_{12}}/{(Y_d)_{12}}= 1$ can well be compatible with $m_\mu/m_s = 9/2$ or $m_\mu/m_s = 6$.

\subsubsection{Charged lepton corrections to neutrino mixing angles in GUT models}
In many GUT models of fermion masses and mixings, characteristic predictions can arise for the neutrino mixing angles which are, however, perturbed by the mixing coming from the charged lepton sector (see e.g.\ \cite{Antusch:2005kw}). One typical example is the leptonic mixing angle $\theta_{13}$. In many models the 1-3 mixing from the neutrino sector is very small or even zero ($\theta_{13}^\nu = 0$). Nevertheless a total lepton mixing $\theta_{13}$ can be induced from the possible corrections caused by mixing in the charged lepton mass matrix and is then given by
\begin{eqnarray}
\theta_{13} \approx \frac{\theta_{12}^e}{\sqrt{2}}  \,,
\end{eqnarray}   
where $\theta_{12}^e$ is the charged lepton 1-2 mixing angle given (for a hierarchical mass matrix) by $\theta_{12}^e \approx (Y_e)^2_{12}/(Y_e)^2_{22}$. Assuming for instance $(Y_e)^2_{12}/(Y_d)^2_{12} = 1$ and 
$|(Y_e)^2_{22}/(Y_d)^2_{22}| \approx m_\mu/m_s = 3$ we obtain $\theta_{13} \approx \theta_{12}^d /(3\sqrt{2})$ where $\theta_{12}^d$ is the 1-2 mixing of the down-type quark mass matrix $Y_d$. Interestingly, in many GUT models $\theta_{12}^d$ is approximately equal to the Cabibbo angle $\theta_C \approx 13^\circ$, which under the above assumptions would yield $\theta_{13} \approx 3^\circ$. This value emerges in many models as prediction for the neutrino mixing $\theta_{13}$, closely related to the GJ relation $m_\mu/m_s = 3$. 

In this context we would like to remark that the alternative GUT predictions $m_\mu/m_s = 9/2$ and $m_\mu/m_s = 6$ can lead to new predictions for the leptonic mixing angle $\theta_{13}$, following the above chain of arguments. In particular, when $m_\mu/m_s = 9/2$ is realised in a unified model it could predict 
\begin{eqnarray}
\theta_{13}\approx 2\theta_C /(9\sqrt{2}) \approx 2^\circ\;.
\end{eqnarray} 
Analogously, $m_\mu/m_s = 6$ could lead to the prediction 
\begin{eqnarray}
\theta_{13}\approx \theta_C /(6\sqrt{2}) \approx 1.5^\circ
\end{eqnarray}
 for the still unmeasured leptonic mixing angle. Additional predictions are possible when the assumption $(Y_e)^2_{12}/(Y_d)^2_{12} = 1$ is replaced by a different group theoretical Clebsch factor.

\section{Summary and conclusions} \label{sec:Conclusions}

GUT predictions for the ratios of quark and lepton masses  can arise after GUT symmetry breaking from group theoretical Clebsch factors and are characteristic properties of unified flavour models. To compare the GUT scale predictions with experimental data, it is crucial to carefully include SUSY threshold corrections. Their effects depend on the low energy SUSY parameters and are particularly relevant for large $\tan \beta$. 

Our study consists of two parts:
  
In the first part (section 2) we have derived possible alternative GUT predictions for the ratios $m_e/m_d$, $m_\mu/m_s$, $y_\tau/y_b$ and $y_t/y_b$ at the unification scale (see tables \ref{tab:SU5messenger} - \ref{tab:PSrelations}). We have assumed a unified gauge group $SO(10)$ which is broken to the MSSM at the GUT scale via the $SU(5)$ or Pati-Salam (PS) breaking chain.

In the second part (section 3), we have analysed which GUT scale ratios are allowed when phenomenological constraints from electroweak precision observable, $B$ physics, $(g-2)_\mu$, mass limits on sparticles from direct searches as well as, optionally, dark matter constraints are taken into account. For explicitness, we have considered the three common minimal SUSY breaking scenarios mAMSB, mGMSB and CMSSM, which provide boundary conditions for the soft SUSY breaking parameters at high energies.

\begin{figure}
 \centering
 \includegraphics[scale=0.4]{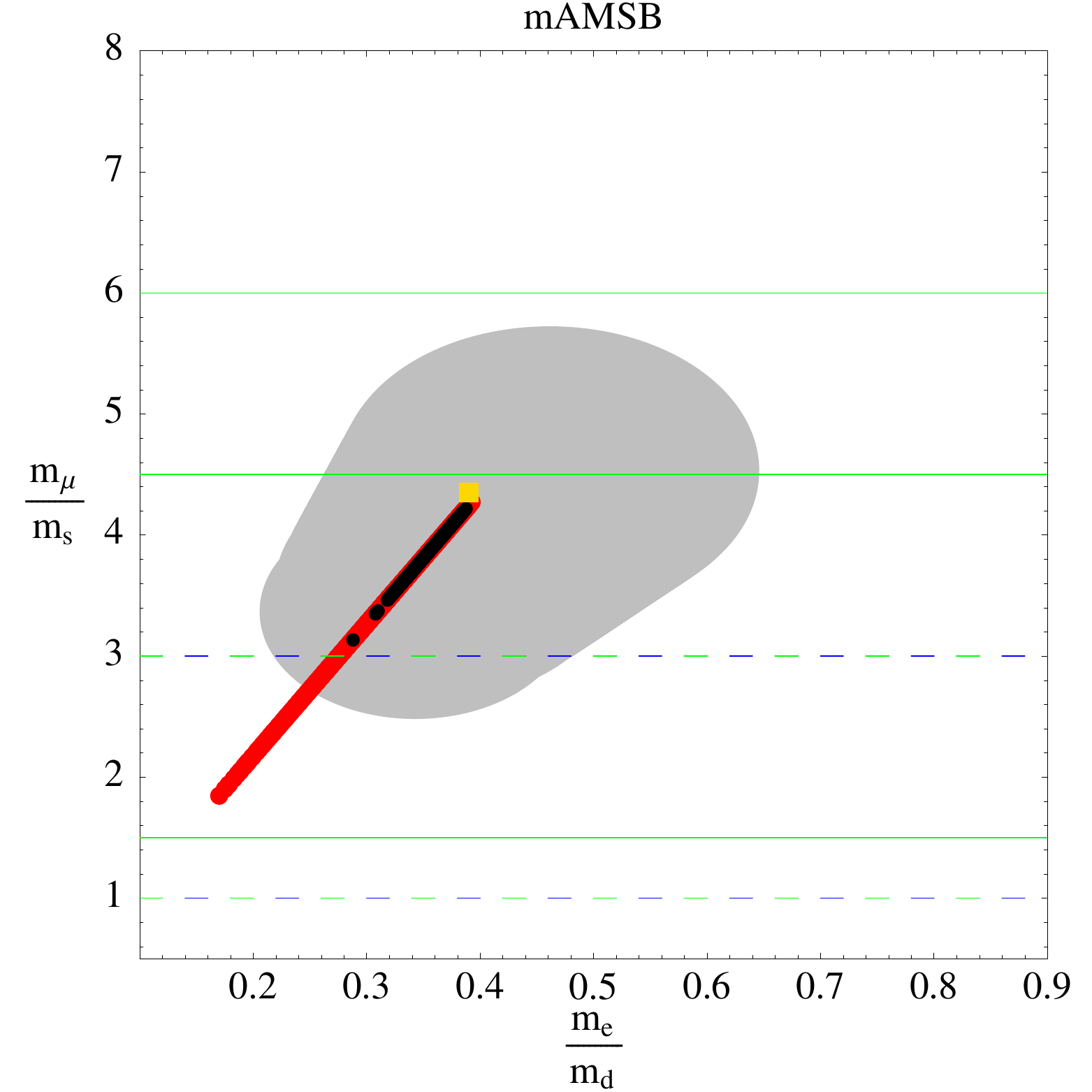}
 \includegraphics[scale=0.4]{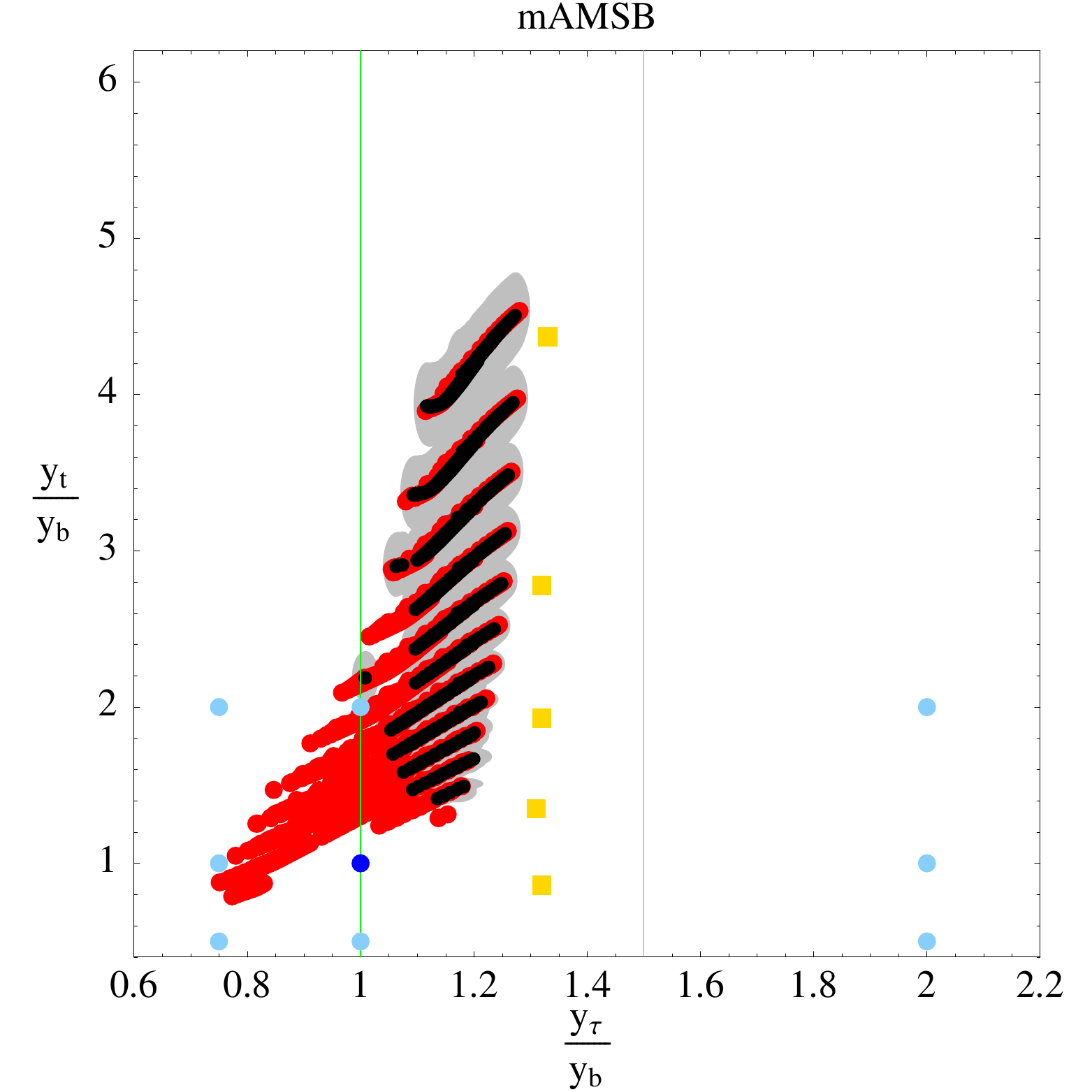}
 \includegraphics[scale=0.4]{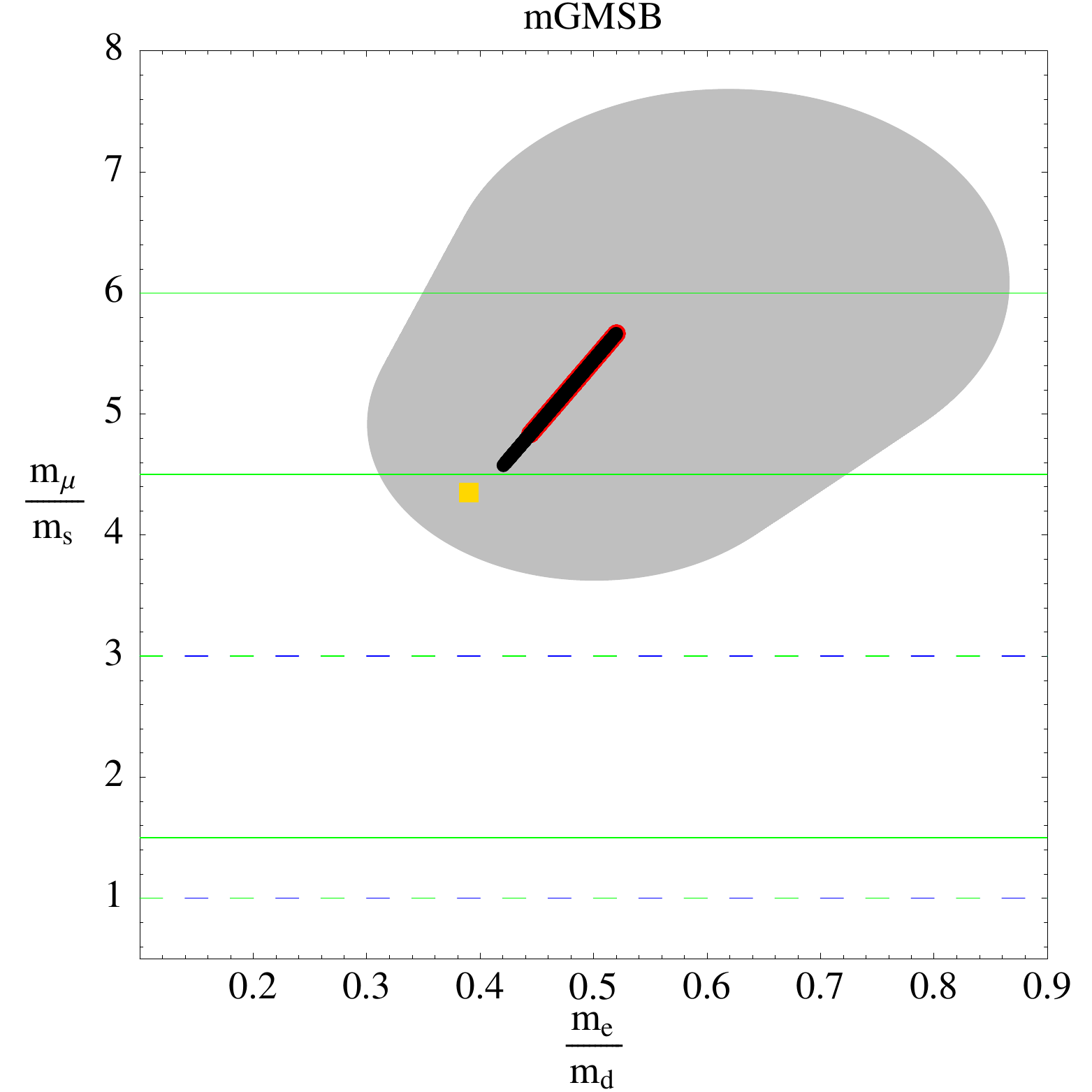}
 \includegraphics[scale=0.4]{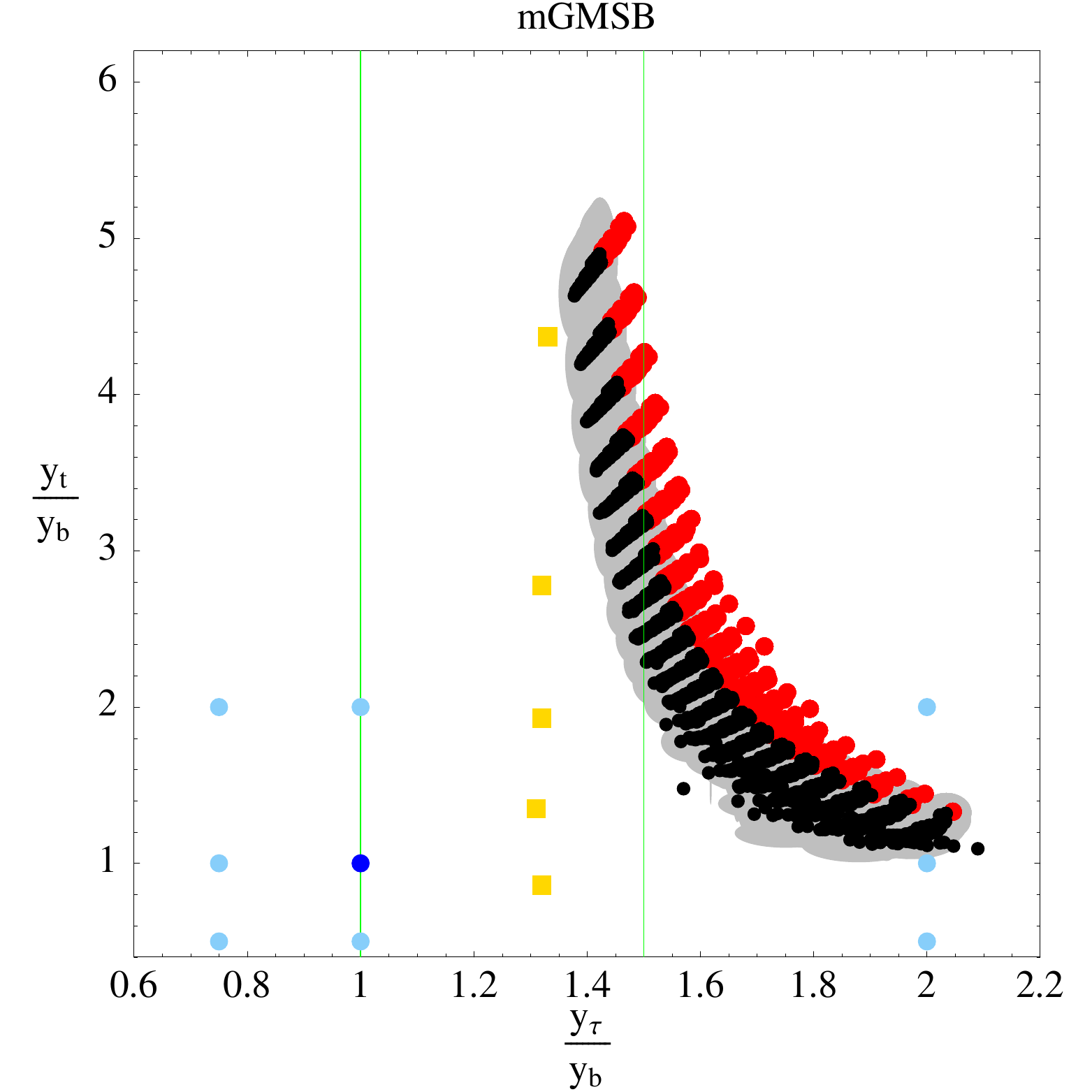}
 \includegraphics[scale=0.4]{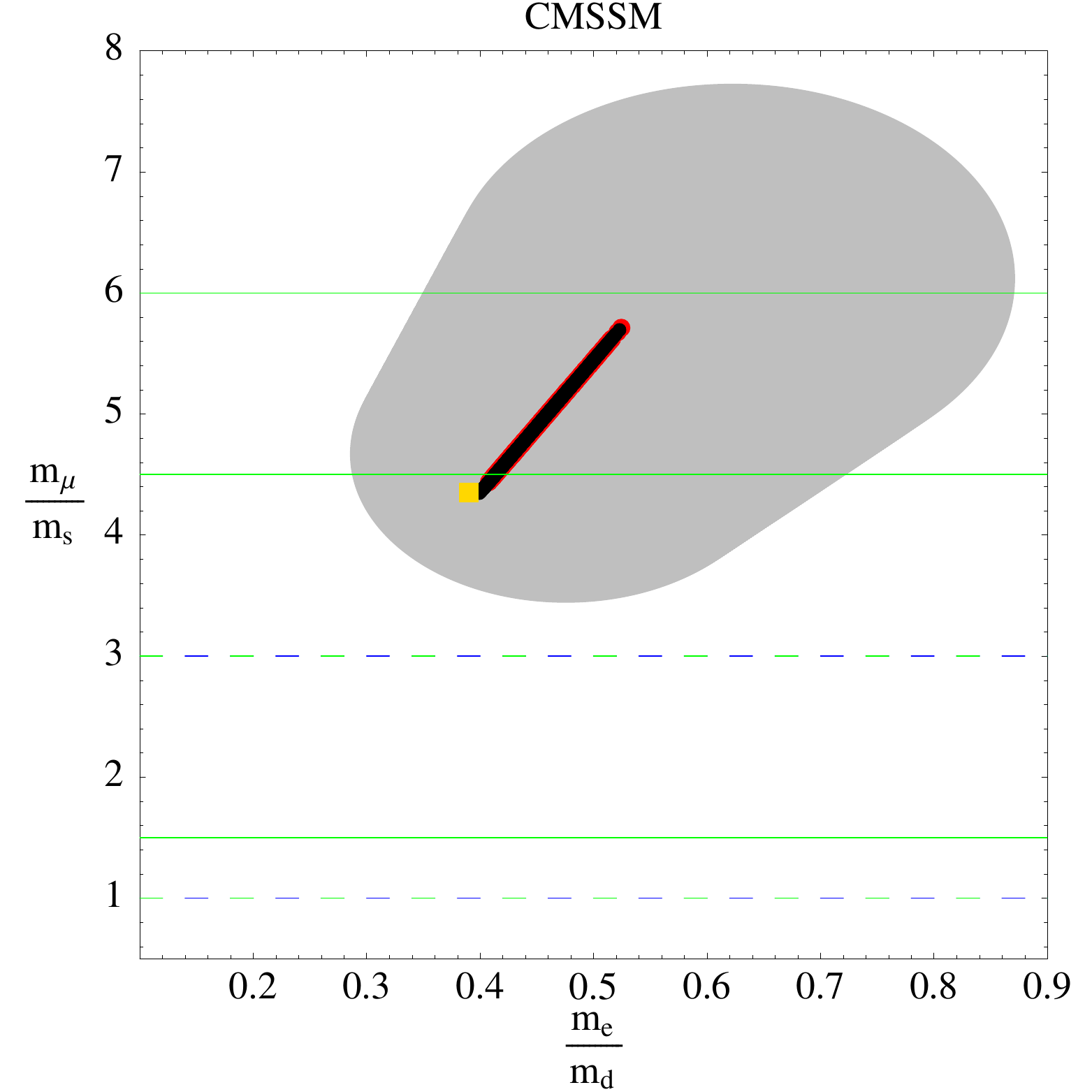}
 \includegraphics[scale=0.4]{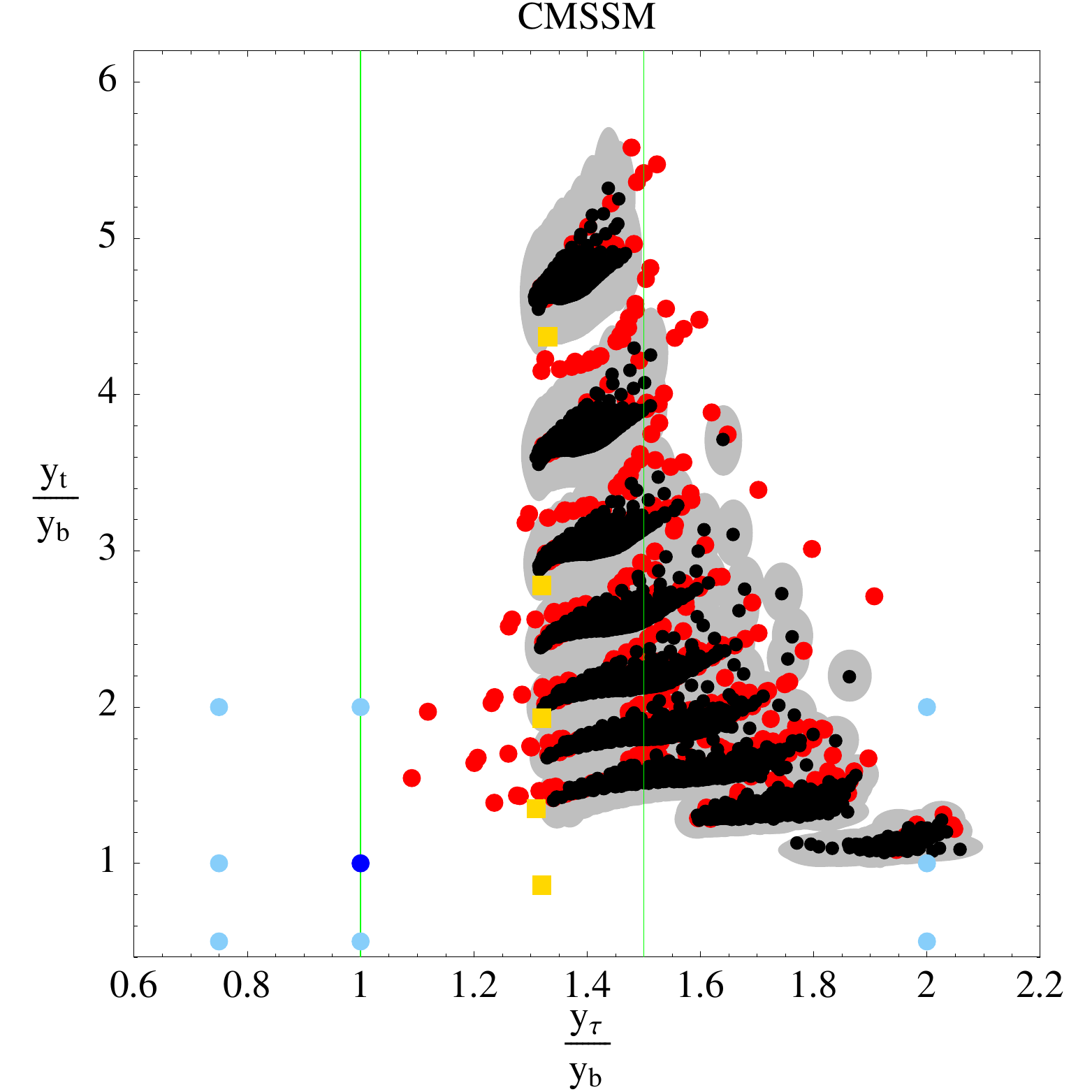}
 \caption{Final results for mAMSB, mGMSB and CMSSM. 
          The (red) black points are the (excluded) allowed points after
          applying the constraints. The grey regions indicate the uncertainties 
	  from experimental quark mass errors. 
	  The green lines are predictions from $SU(5)$, the dashed lines from $SU(5)$ and PS 
	  and the (light) blue points from PS (dimension-six operators). 
	  The yellow squads are the GUT scale Yukawa ratios without including
          SUSY threshold corrections for
          $\tan \beta = 20$, $30$, $40$, $50$ and $60$ from top to bottom.
   \label{fig:finalresults}}
\end{figure}

\begin{figure}
 \centering
 \includegraphics[scale=0.4]{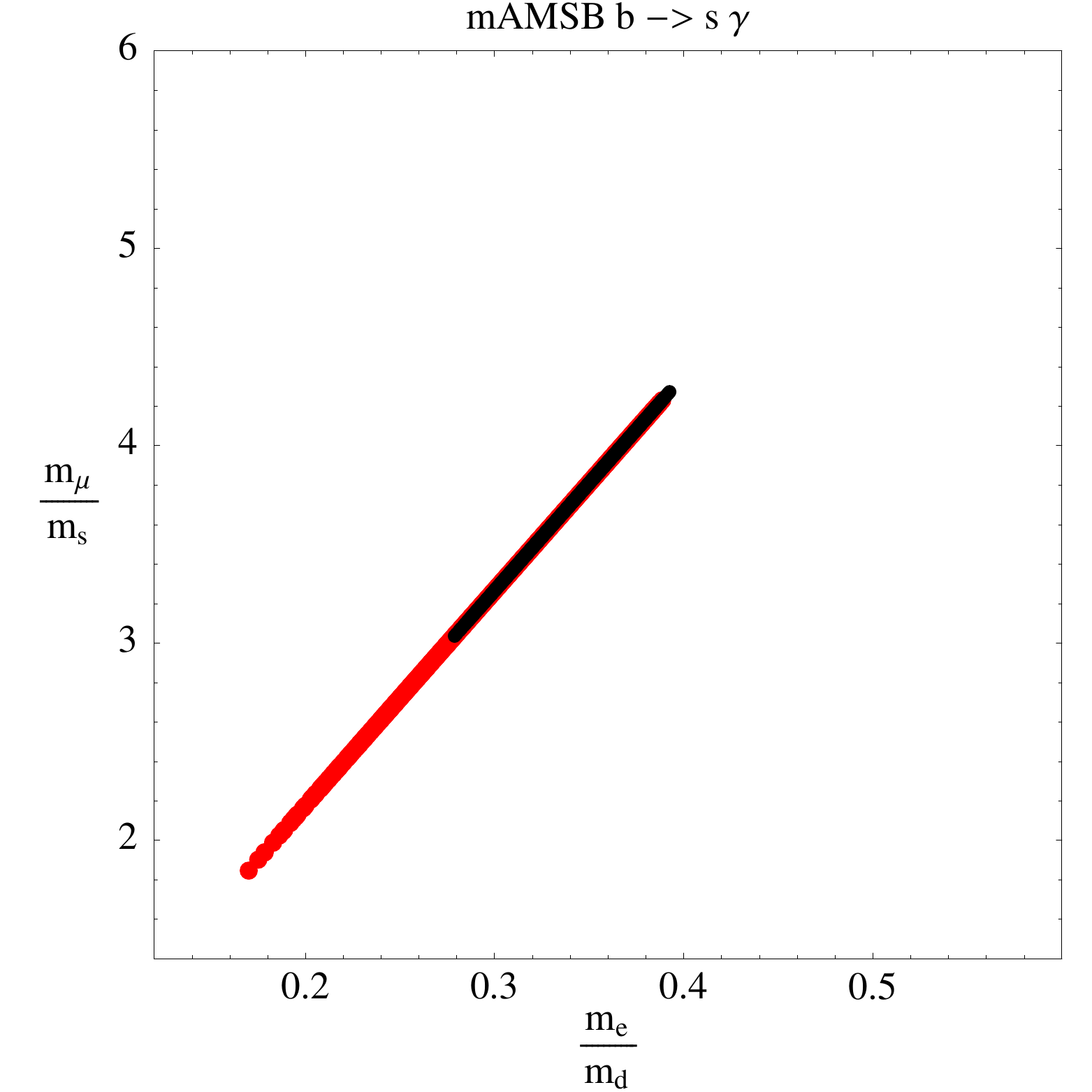}
 \includegraphics[scale=0.4]{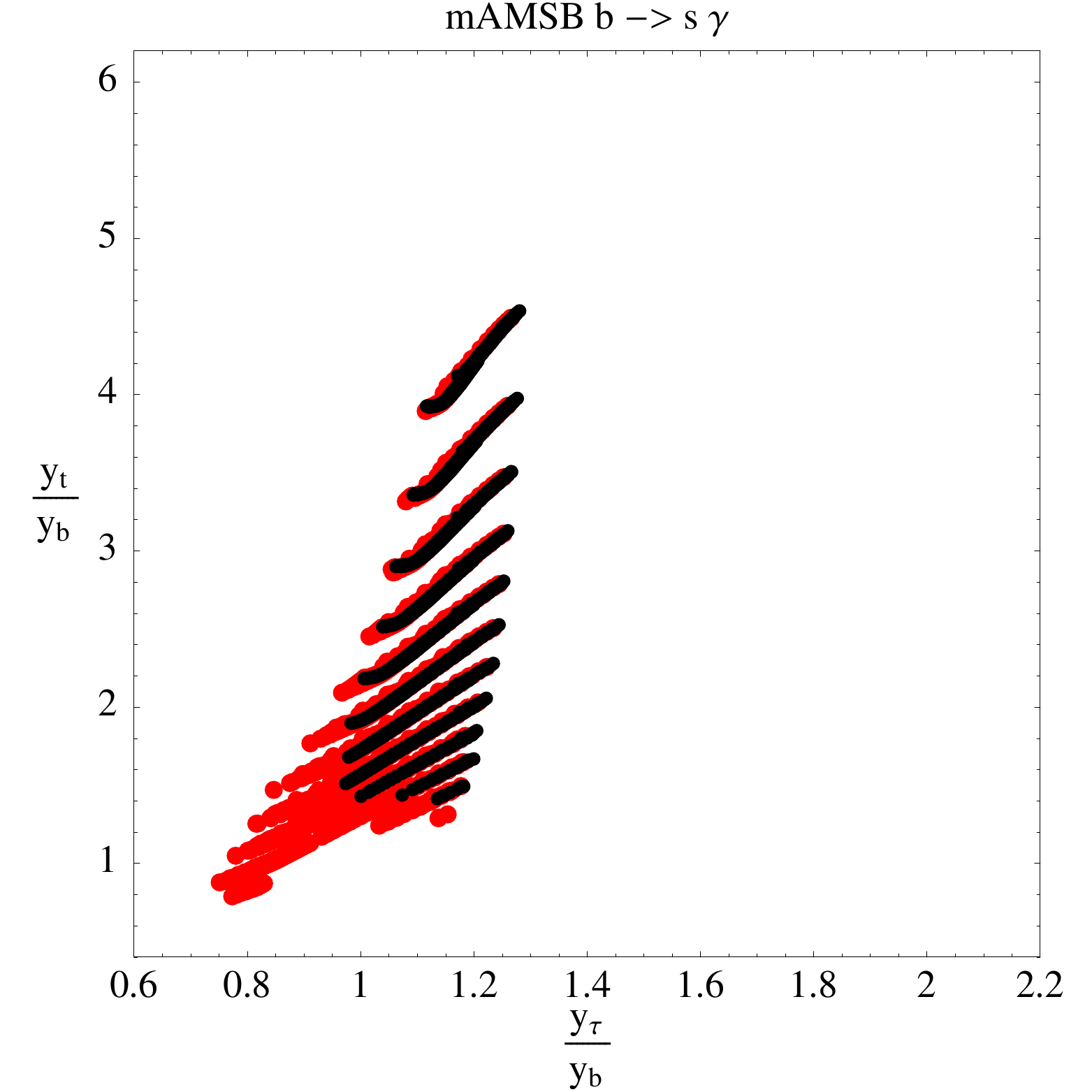}
 \includegraphics[scale=0.4]{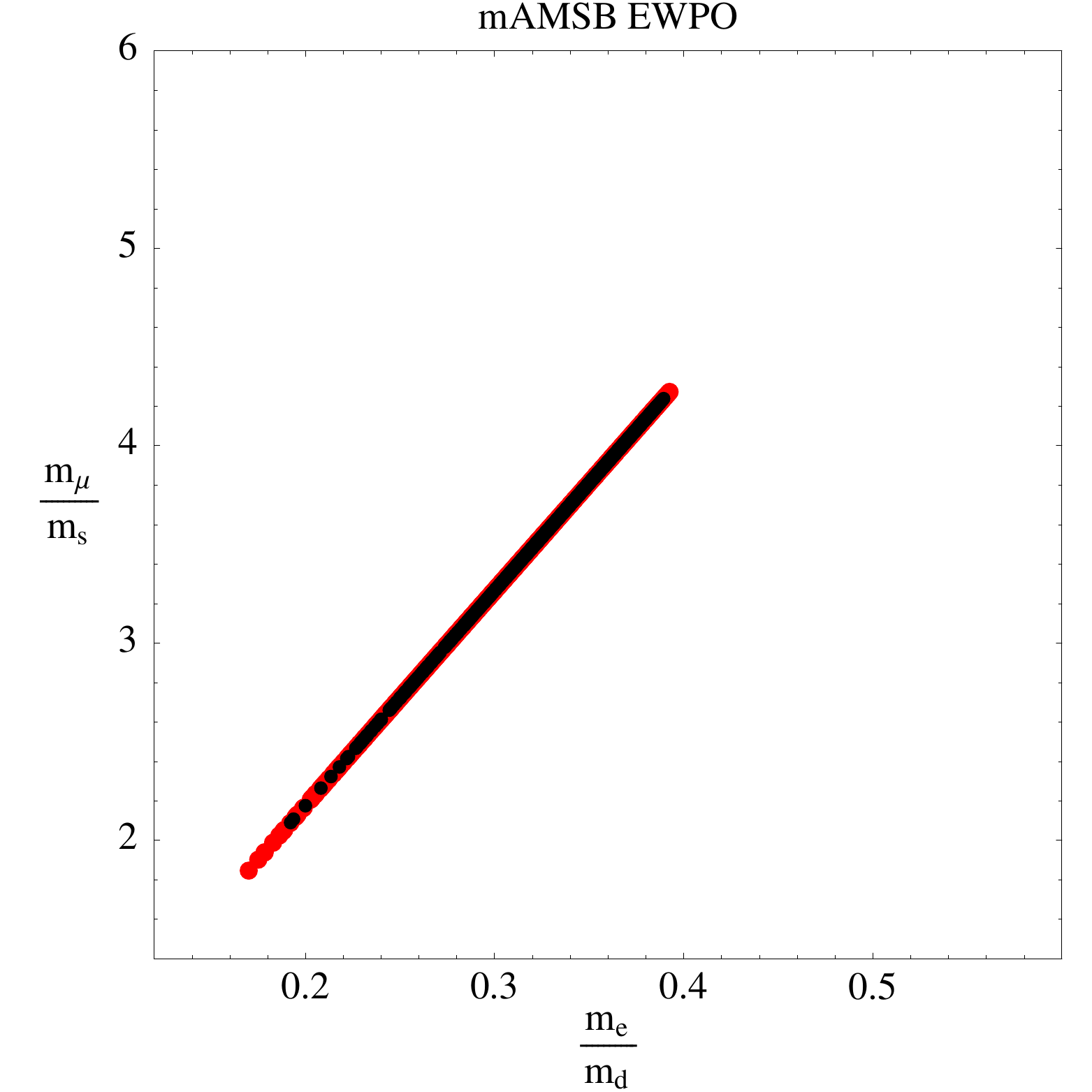}
 \includegraphics[scale=0.4]{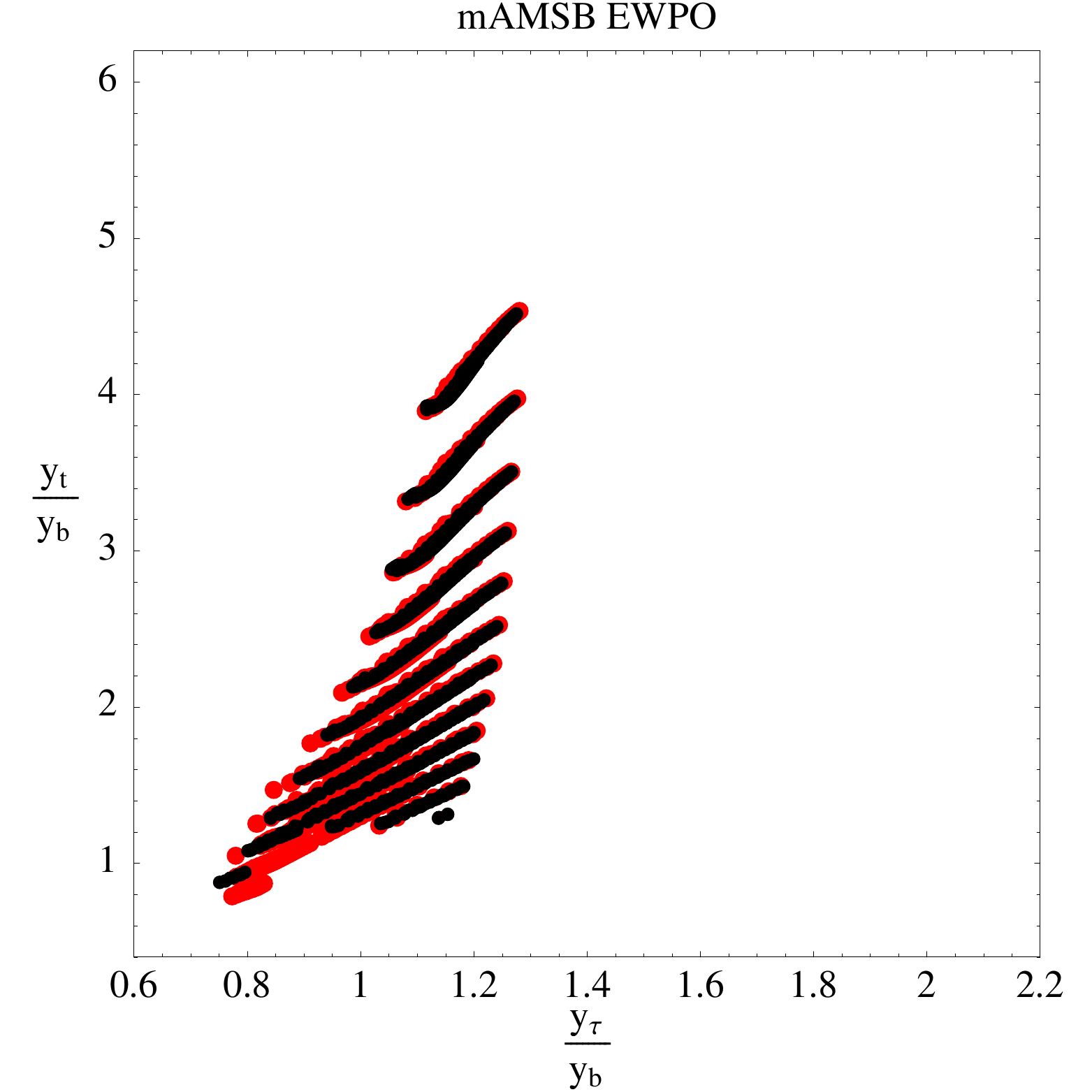}
 \includegraphics[scale=0.4]{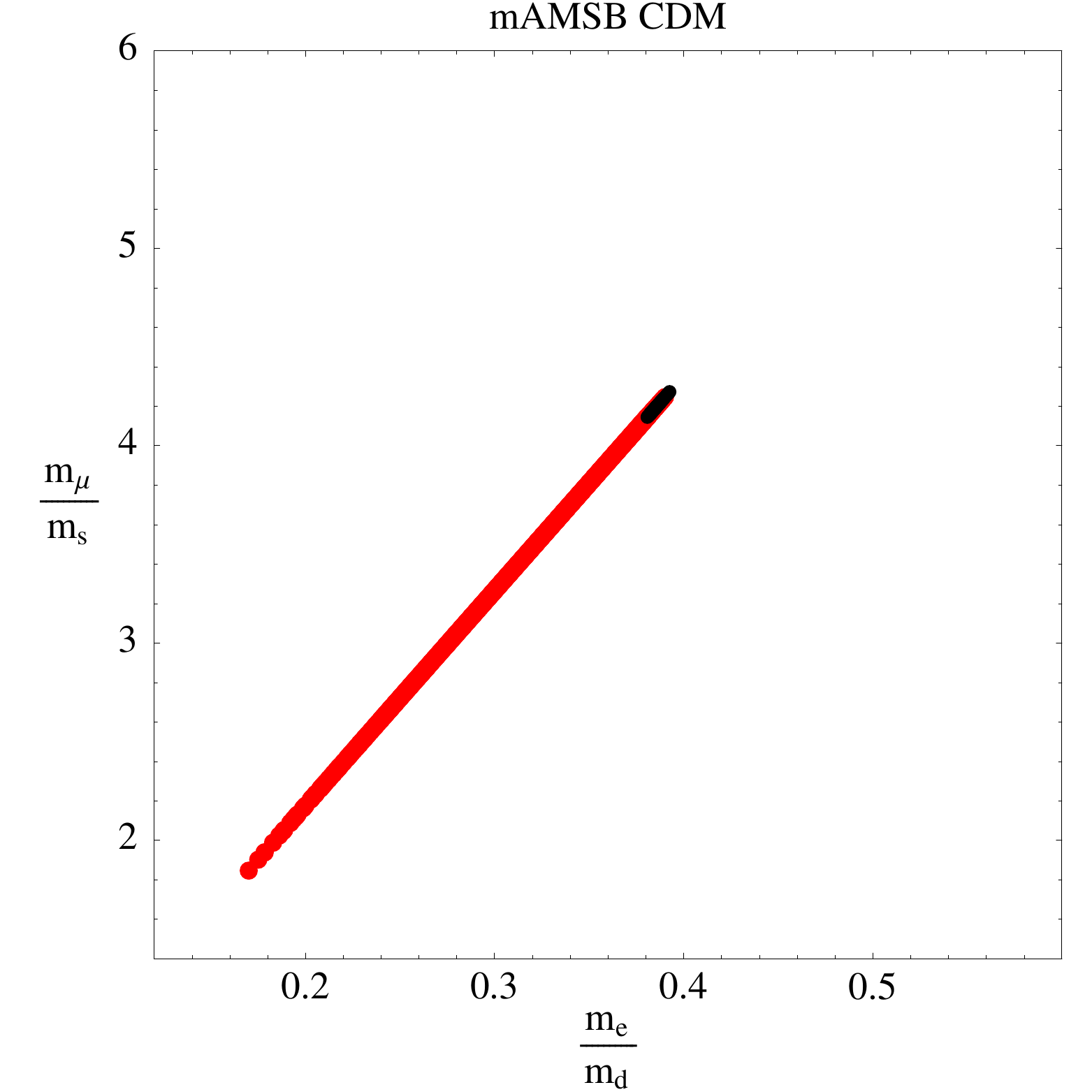}
 \includegraphics[scale=0.4]{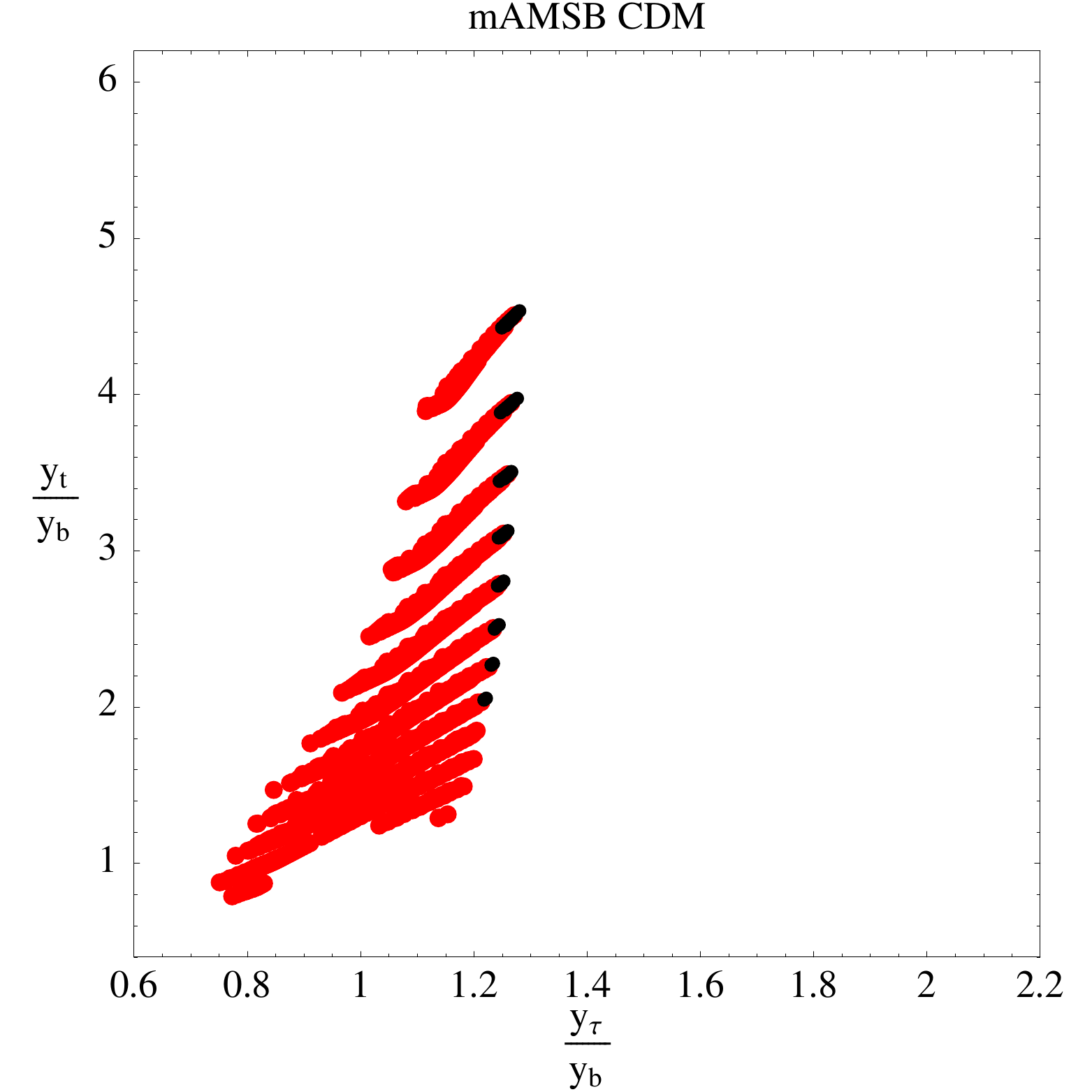}
 \caption{Impact of the constraints from  $b \rightarrow s \gamma$,
          electroweak precision observables (EWPO) as well as from
          dark matter in mAMSB (c.f.\ section 3.4.1). 
	  The latter criterion is not used as a constraint
          for the final results in figure \ref{fig:finalresults}. 
	  Red points denote parameter points which are excluded by the constraint 
	  while black dots indicate parameter points which are allowed. 
	  In the plots on the right the different lines of
          points correspond to different values of $\tan \beta$,
          increasing from top to bottom. 
          \label{fig:plots_mAMSB}}
\end{figure}

\begin{figure}
 \centering
 \includegraphics[scale=0.4]{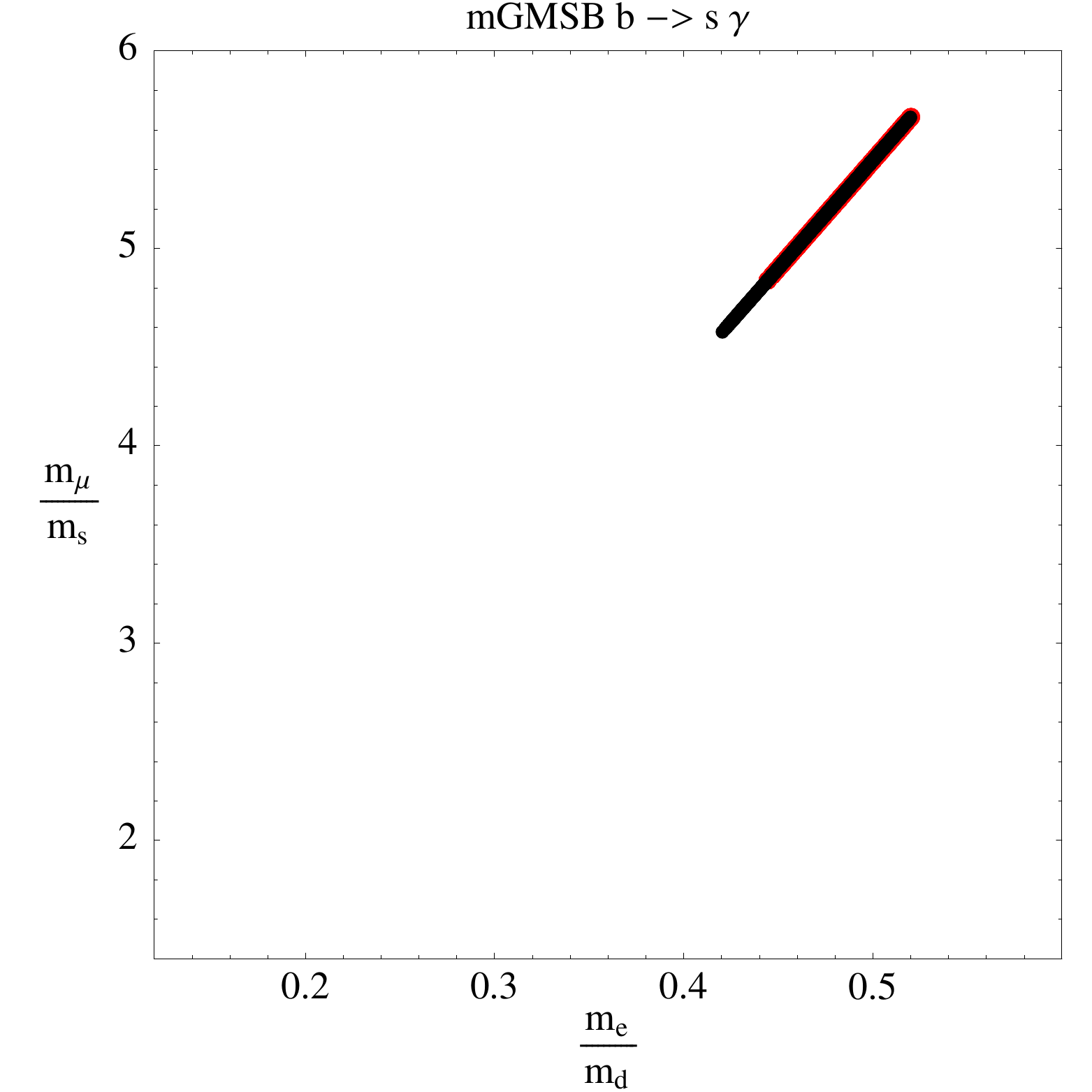}
 \includegraphics[scale=0.4]{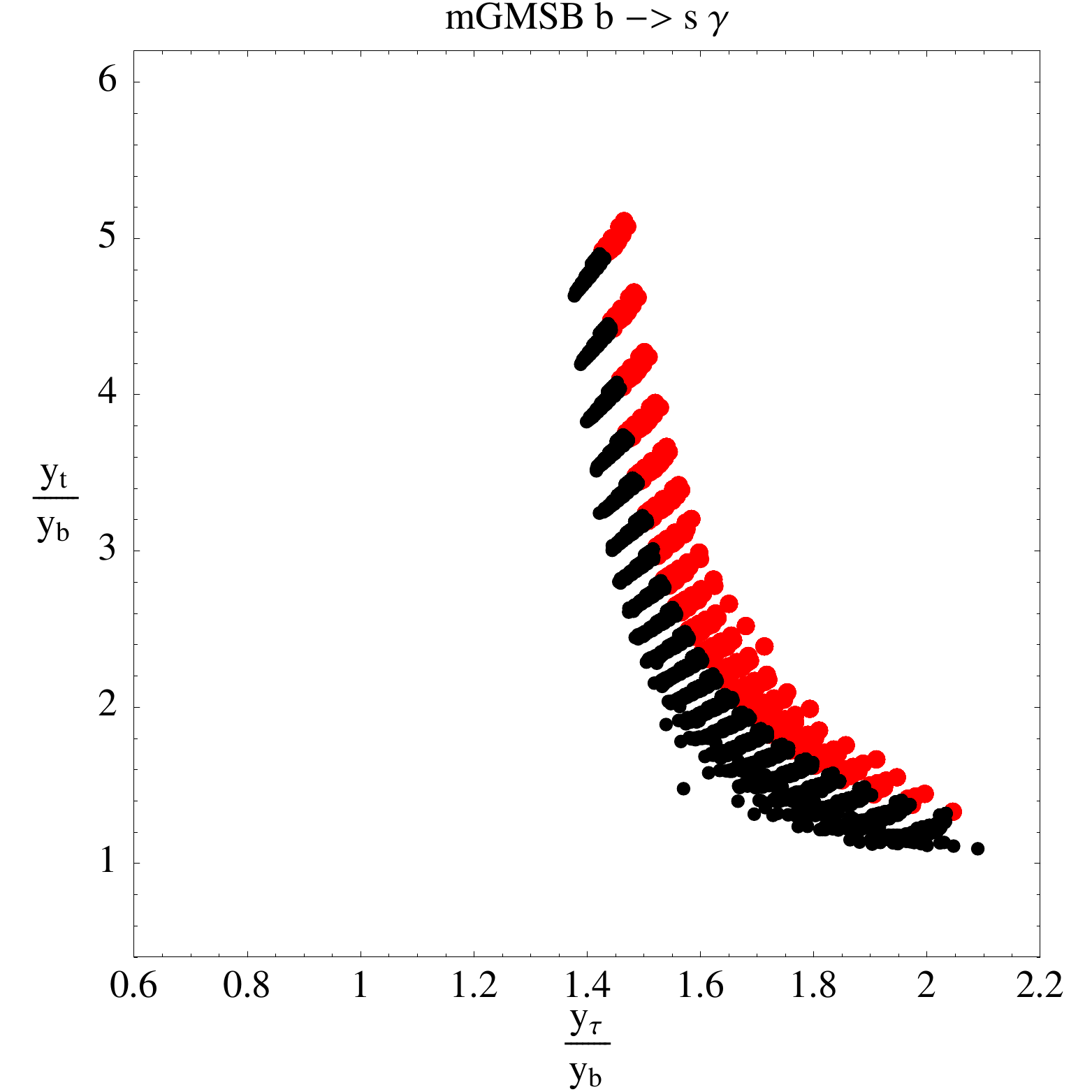}
 \includegraphics[scale=0.4]{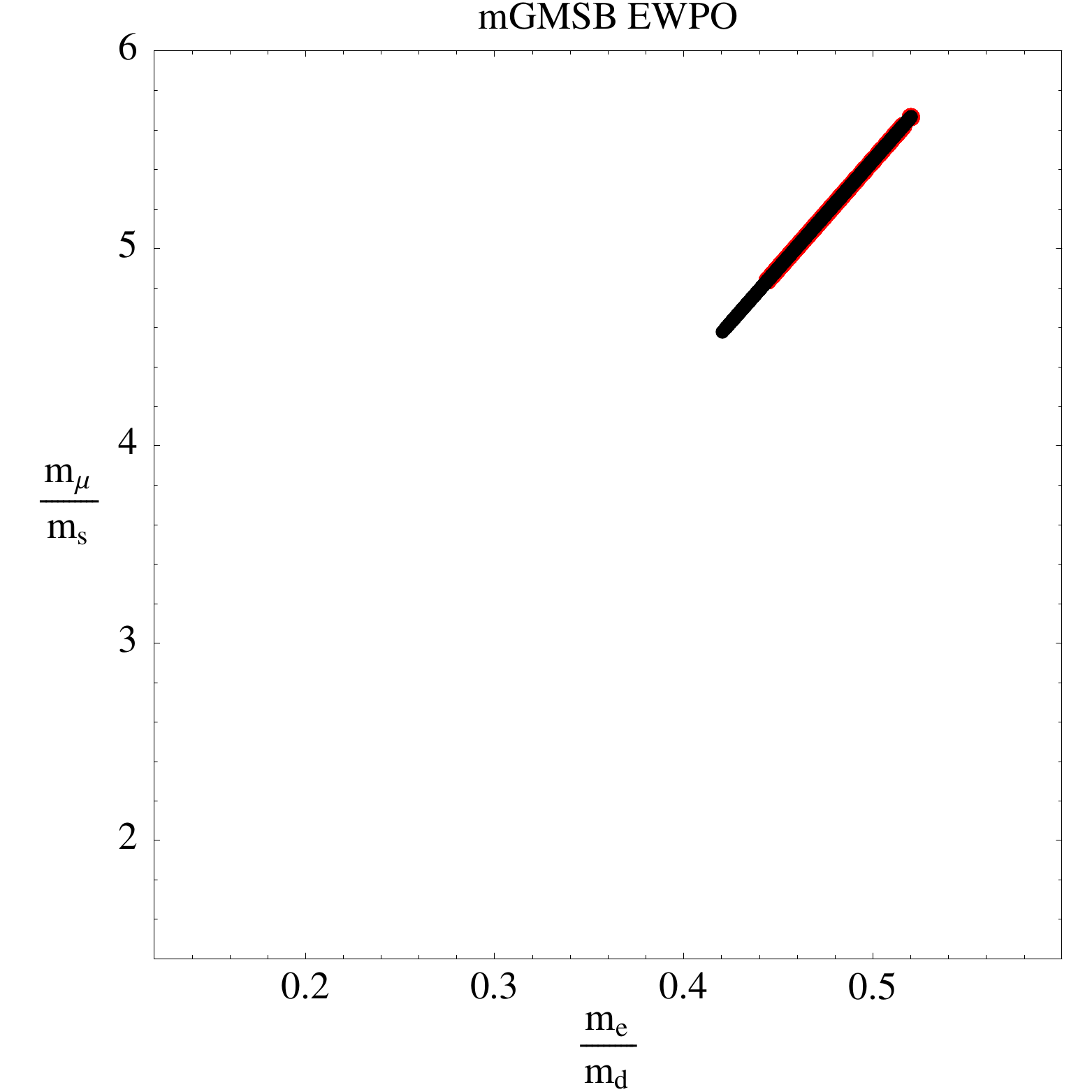}
 \includegraphics[scale=0.4]{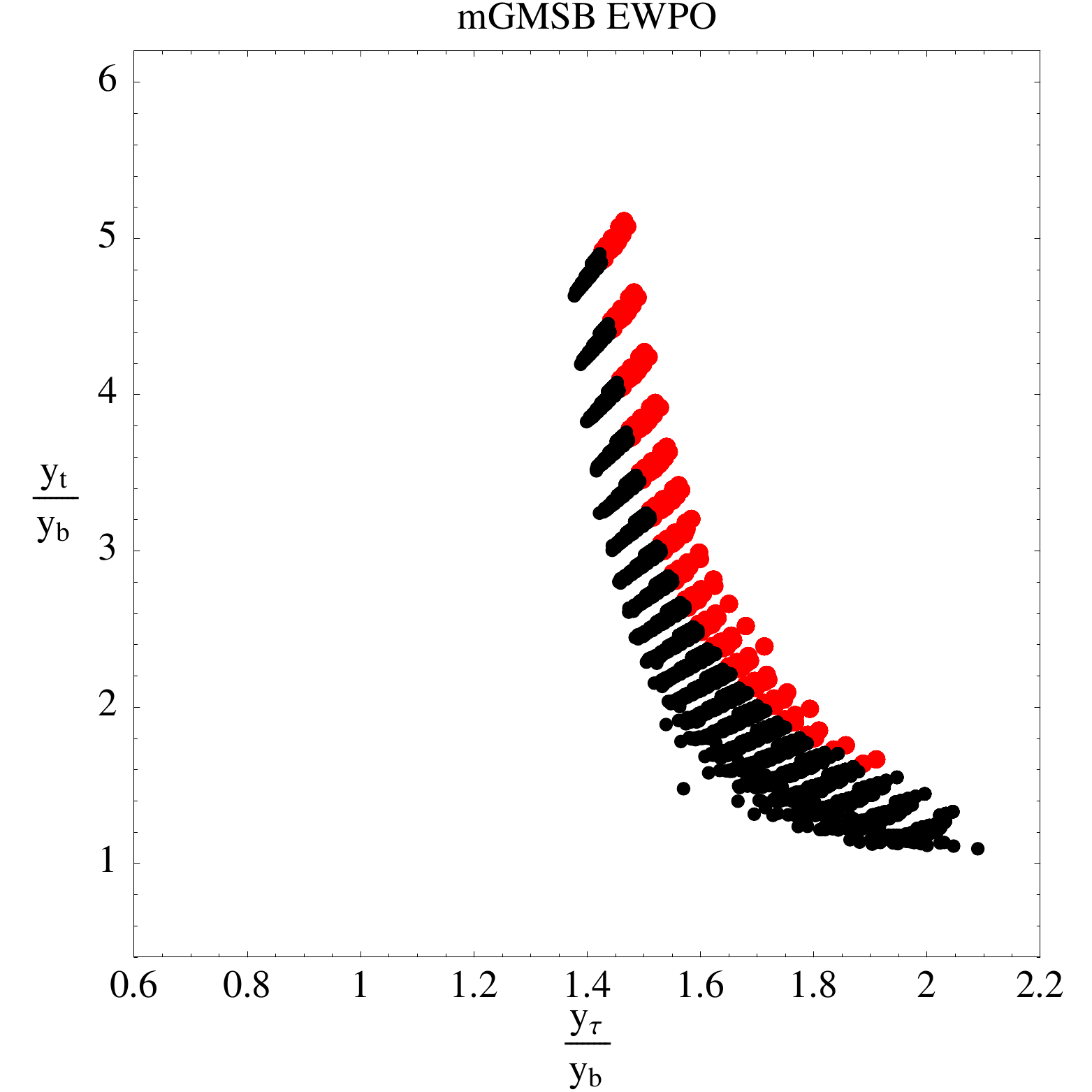}
 \caption{Impact of the constraints from  $b \rightarrow s \gamma$ and 
          electroweak precision observables (EWPO) in mGMSB (c.f.\ section 3.4.2). 	  
	  Red points denote parameter points which are excluded by the constraint 
	  while black dots indicate parameter points which are allowed. 
	  In the plots on the right the different lines of
          points correspond to different values of $\tan \beta$,
          increasing from top to bottom. 
          \label{fig:plots_mGMSB} }
\end{figure}

\begin{figure}
 \centering
 \includegraphics[scale=0.4]{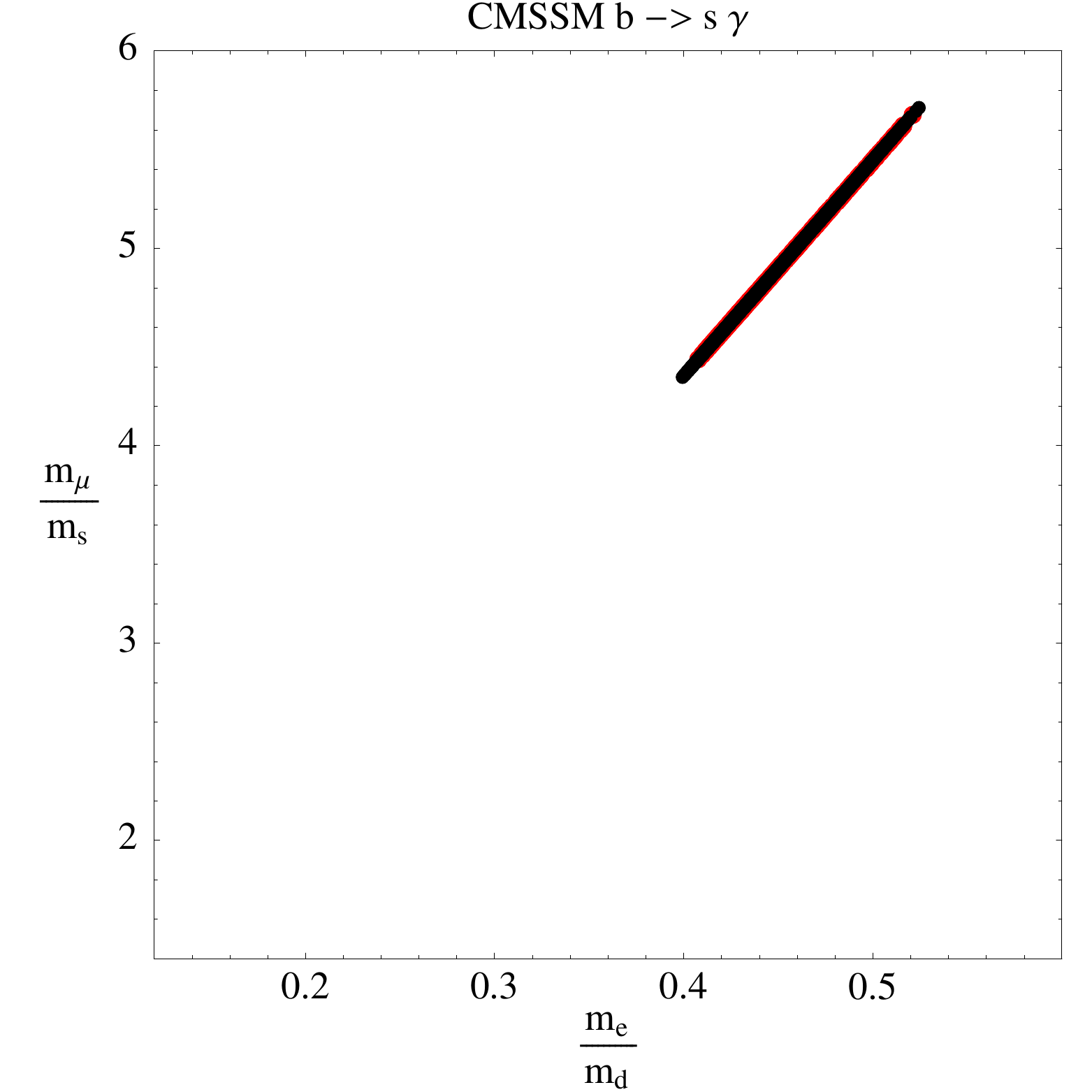}
 \includegraphics[scale=0.4]{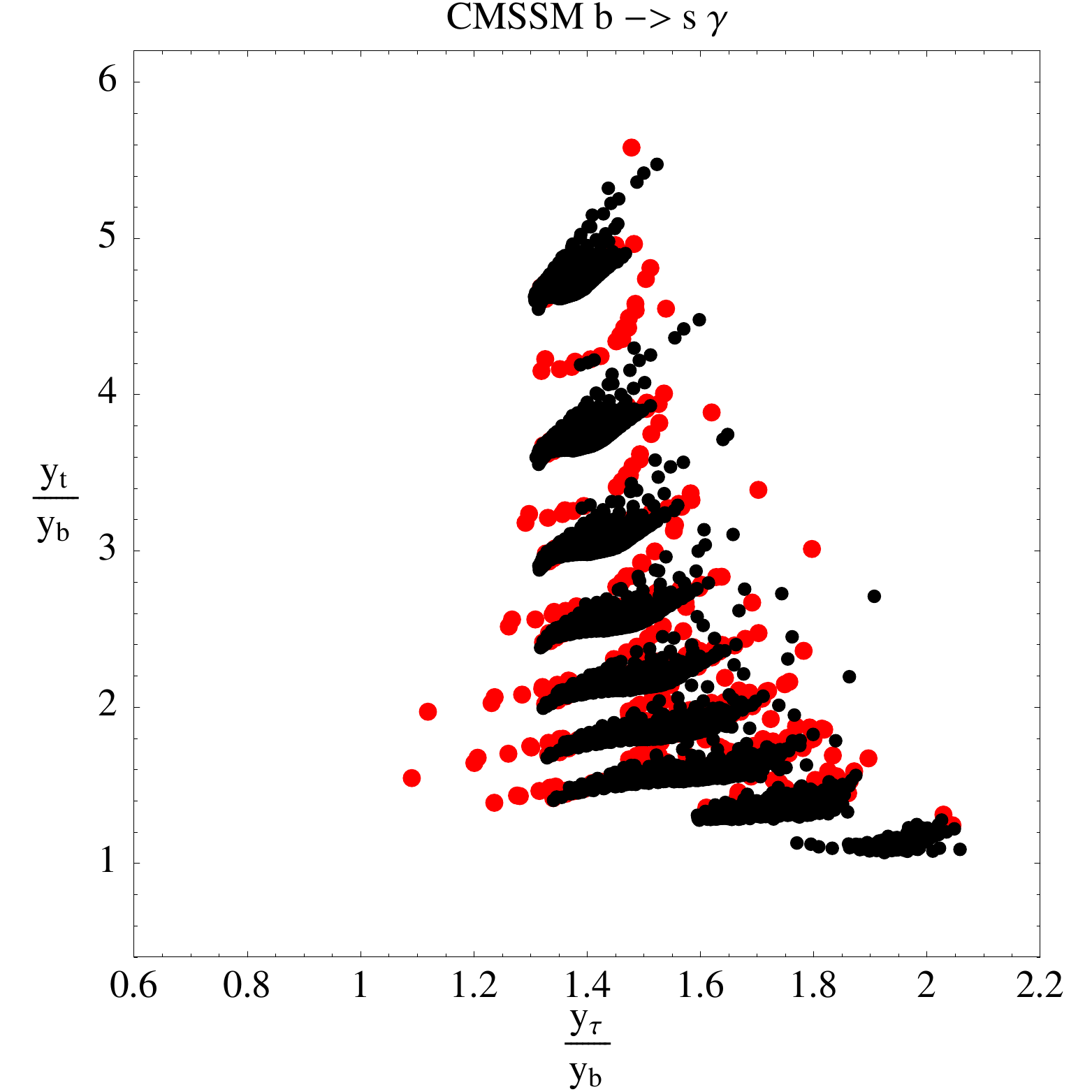}
 \includegraphics[scale=0.4]{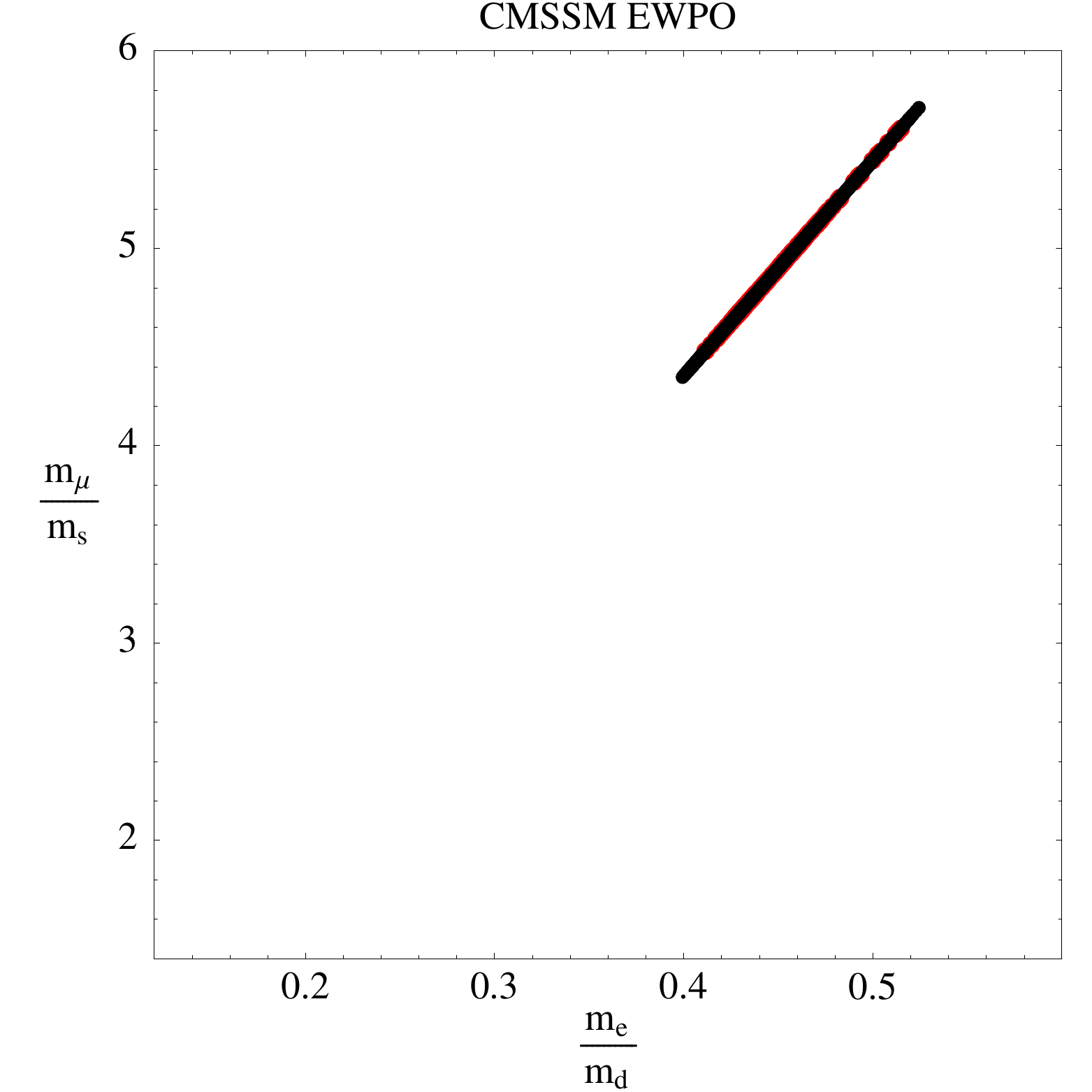}
 \includegraphics[scale=0.4]{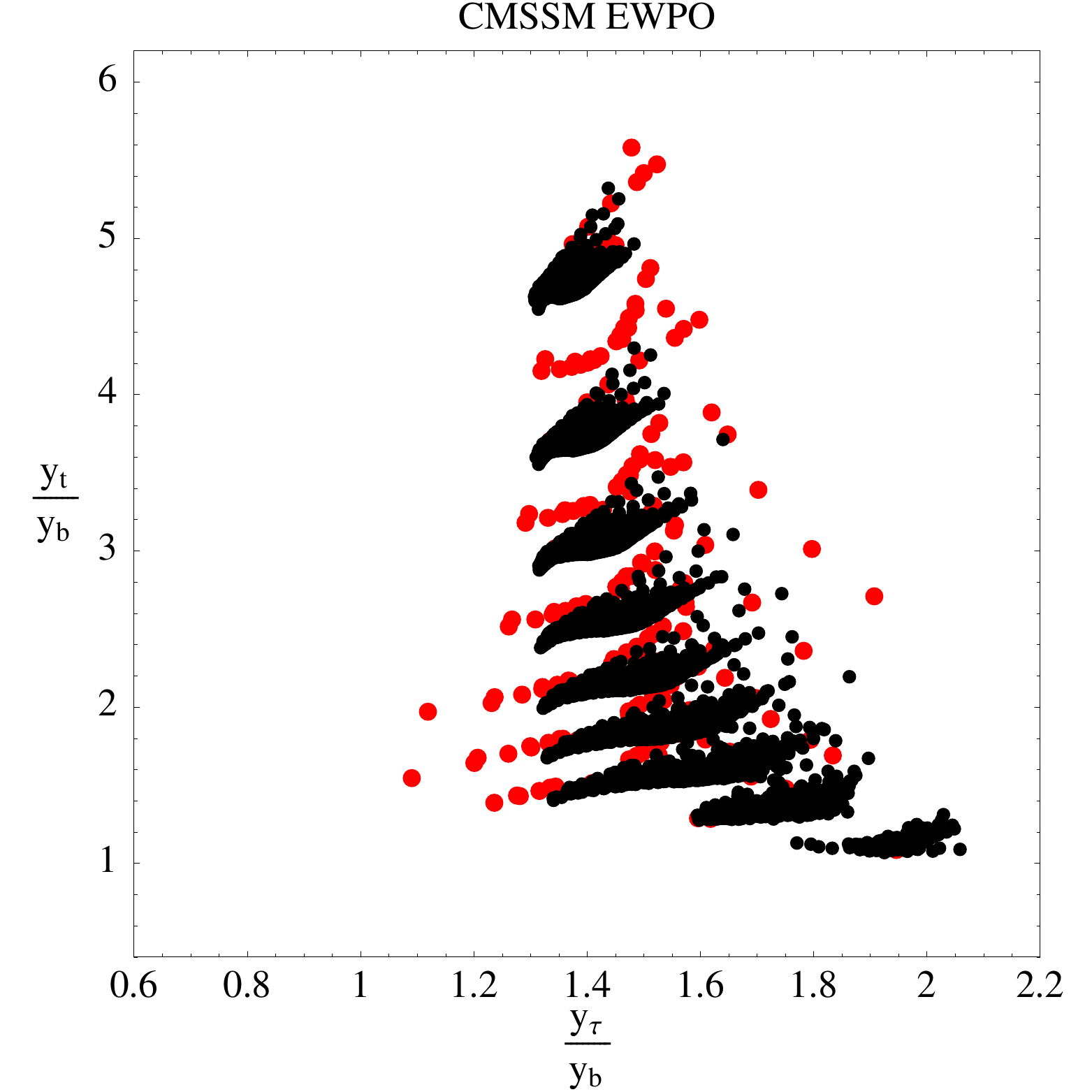}
 \includegraphics[scale=0.4]{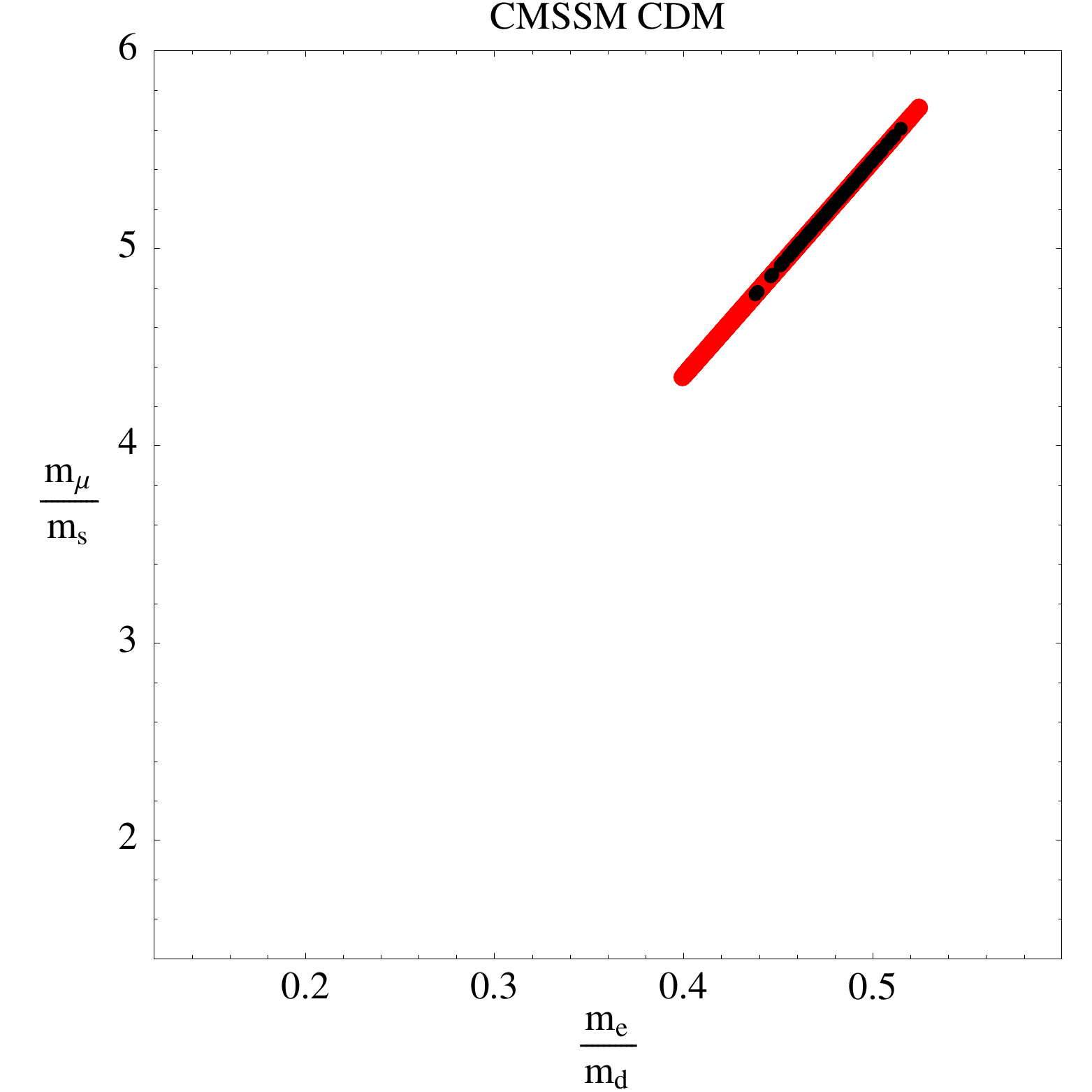}
 \includegraphics[scale=0.4]{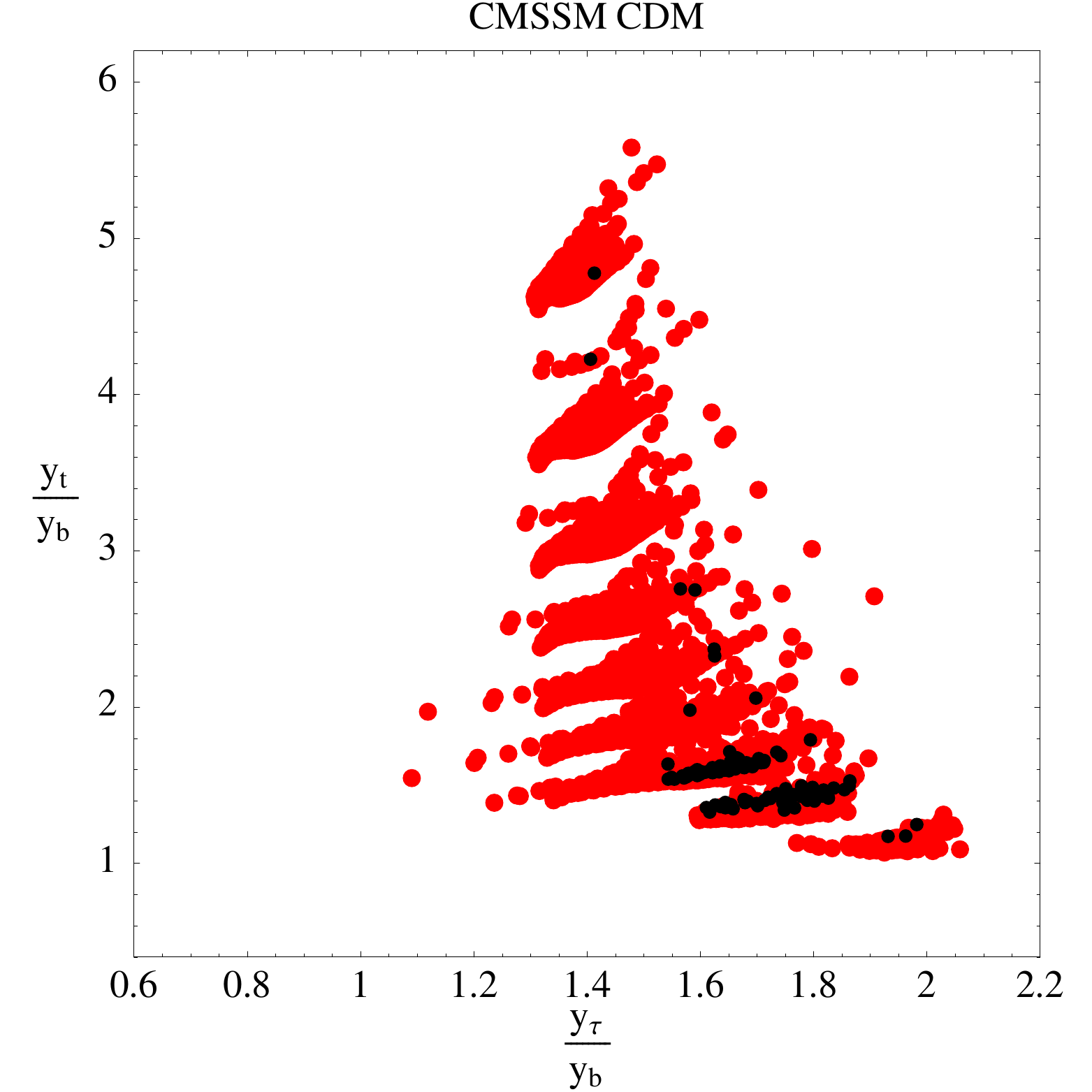}
 \caption{Impact of the constraints from  $b \rightarrow s \gamma$,
          electroweak precision observables (EWPO) as well as from
          dark matter in CMSSM (c.f.\ section 3.4.3). 
	  The latter criterion is not used as a constraint
          for the final results in figure \ref{fig:finalresults}. 
	  Red points denote parameter points which are excluded by the constraint 
	  while black dots indicate parameter points which are allowed. 
	  In the plots on the right the different lines of
          points correspond to different values of $\tan \beta$,
          increasing from top to bottom. 
          \label{fig:plots_CMSSM} }
\end{figure}

From comparing the GUT scale predictions with the phenomenologically allowed ranges within mAMSB, mGMSB and CMSSM (see figure \ref{fig:finalresults}), we have obtained the following main results (c.f.\ section 4):  
\begin{itemize}

\item The Georgi-Jarlskog (GJ) relation of $m_\mu/m_s = 3$ at $M_{GUT}$ is incompatible with mGMSB and CMSSM, however it can be realised in mAMSB for intermediate $\tan \beta$ ($\sim 30$) and relatively heavy sparticle spectrum. While the possibility of $m_\mu/m_s = 3$ in AMSB-like SUSY breaking scenarios has been suggested already in \cite{Ross:2007az,Antusch:2008tf}, our results show that the realisation of $m_\mu/m_s = 3$ can be consistent  with phenomenological constraints.

\item Regarding alternative predictions for $m_\mu/m_s$, we find that in mGMSB and CMSSM, $m_\mu/m_s = 9/2$ or $m_\mu/m_s = 6$ are possible, where the former corresponds to small threshold effects and small or moderate $\tan \beta$ whereas the latter corresponds to large threshold corrections and large $\tan \beta$. In mAMSB with small or moderate $\tan \beta$, $m_\mu/m_s = 9/2$ is also consistent. Both predictions, $m_\mu/m_s = 9/2$ and $m_\mu/m_s = 6$, can be realised in unified theories based on $SU(5)$ (or on $SO(10)$ with breaking chain via $SU(5)$). Smaller predictions such as $m_\mu/m_s = 2$ proposed in \cite{Antusch:2005ca} are phenomenologically disfavoured in all three scenarios. 

\item In the considered scenarios we found no example where third family Yukawa unification $y_t = y_b = y_\tau$ was realised. Interestingly, even in mAMSB we did not find any consistent parameter point, in contrast to the claims in \cite{Ross:2007az,Antusch:2008tf}, due to inconsistencies with tachyons, EWPO and B-physics observables. 
However, we would like to remark that mAMSB is only a minimal scenario and $y_t = y_b = y_\tau$ may in principle be allowed in different models with anomaly mediation. On the other hand, our results suggest that it might be difficult to realise such large threshold effects in a phenomenological consistent way. 
In the CMSSM (as well as in mGMSB) with $\mbox{sgn}(\mu M_3)$ positive, the threshold corrections generically enlarge $y_\tau / y_b$ such that third family Yukawa unification is not allowed. However under certain conditions in CMSSM, in particular with large negative trilinear coupling $A_t$, one can in principle find tuned regions with $y_t = y_b = y_\tau$, which are however excluded by the experimental constraints as argued  in \cite{Altmannshofer:2008vr}. 
In figure \ref{fig:finalresults} there are only a few excluded points close to $y_t = y_b = y_\tau$, which is due to the fact that we have not tuned any parameters for our scan.

\item There are alternative relations between the third generation Yukawa couplings $y_t$,  $y_b$ and $y_\tau$ which seem to be favoured compared to third family Yukawa unification: For instance, dimension-six operators in PS can lead to the relation $y_t = 2 y_b = 2 y_\tau$ which is allowed in mAMSB (with intermediate $\tan \beta$ and comparatively heavy SUSY spectrum) or to  $2 y_t = 2 y_b = y_\tau$ which is allowed in mGMSB and CMSSM (with large $\tan \beta$). In mGMSB and CMSSM the relation $y_\tau/y_b = 3/2$ can be realised for moderate values of $\tan \beta$. 

\item It is also interesting to remark that in mAMSB, the GUT predictions $y_t = 2 y_b = 2 y_\tau$ and $m_\mu/m_s = 3$ can be valid for the same region of parameter space. In mGMSB and CMSSM, $y_\tau/y_b = 3/2$ and $m_\mu/m_s = 9/2$ can be realised simultaneously as well as $2 y_t = 2 y_b = y_\tau$ and $m_\mu/m_s = 6$.

\item Furthermore, bounds from thermal overproduction of dark matter may be considered as constraints on the SUSY parameters under the additional assumptions of a stable lightest neutralino and of a ``standard'' cosmological history. These constraints (which are not included in our results shown in figure \ref{fig:finalresults}) would exclude a large range of possible GUT scale values, in particular in mAMSB where only $m_\mu/m_s = 9/2$ would remain as a viable GUT prediction. In the CMSSM, the dark matter bounds are less restrictive (c.f.\ discussion in section 3.4.3) and the relations $y_\tau/y_b = 3/2$, $y_\tau/y_b = 2$, $m_\mu/m_s = 9/2$ and $m_\mu/m_s = 6$ remain allowed.  
\end{itemize}

In summary, we have derived possible new predictions for the GUT scale mass (or Yukawa coupling) ratios $m_\mu/m_s$, $y_\tau/y_b$ and $y_t/y_b$ and confronted them with phenomenological constraints. The soft SUSY breaking scenarios mAMSB, mGMSB and CMSSM have been taken as explicit examples, however our results may hold true approximately in variants of these schemes. The allowed GUT scale ranges for $m_\mu/m_s$, $y_\tau/y_b$ and $y_t/y_b$ have been calculated and compared to the theory predictions.  We found that new GUT scale predictions such as $m_\mu/m_s = 9/2$ or $6$ and $y_\tau/y_b = 3/2$ or $2$ are often favoured compared to the ubiquitous $m_\mu/m_s = 3$ or $y_\tau/y_b =1$. In general, GUT predictions for quark and lepton mass ratios point to characteristic SUSY spectra and breaking mechanisms which can be tested at the CERN LHC and future colliders.

\section*{Acknowledgements}
The authors would like to thank Lorenzo Calibbi for discussions 
and comments on the draft. This work is partially supported by 
the DFG cluster of excellence ``Origin and Structure of the Universe''.

\providecommand{\bysame}{\leavevmode\hbox to3em{\hrulefill}\thinspace}

\end{document}